\documentclass{aa}
\usepackage[varg]{txfonts}   
\usepackage{graphicx}
\usepackage{natbib}

\begin{document}
\title{Calibrating Milky Way dust extinction using cosmological sources}
\author{Edvard M\"ortsell\inst{1,2}}
\institute{Department of Physics, Stockholm University, AlbaNova
        University Center, SE 106 91 Stockholm, Sweden 
        \and The Oskar Klein Centre for Cosmoparticle Physics, 
        Stockholm University, SE 106 91 Stockholm, Sweden\\
        \email{edvard@fysik.su.se}}
\date{Received / Accepted}

\abstract
{}
{To constrain the light extinction properties of Milky Way dust.}
{We investigated the correlations between dust column
density as inferred from infrared data and the observed colours of
celestial objects at cosmological distances with low levels of colour
dispersion.}
{Results derived using colours of quasars, brightest central galaxies,
  and luminous red galaxies are broadly consistent, indicating
  a proportionality constant between the reddening $E(B-V)\equiv
  A_B-A_V$ and the dust column density $D^T$ (given in units of
  MJy/sr) of $p=E(B-V)/D^T\sim 0.02$ and a reddening parameter
  $R_V\equiv A_V/E(B-V)\sim 3$ with fractional uncertainties of
  approximately $10\,\%$.  The data do not provide any evidence for
  spatial variations in the dust properties, except for a possible
  hint of scatter in the dust extinction properties at the longest
  optical wavelengths.}
{}


\maketitle

\section{Introduction}\label{sec:introduction}
\begin{figure}[!htb]
\resizebox{\hsize}{!}{\includegraphics{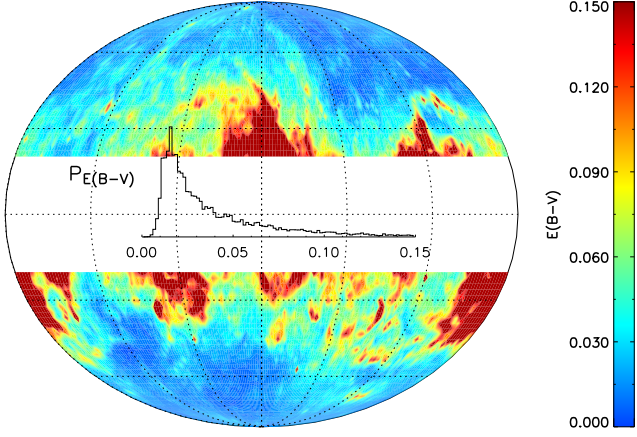}}
\caption{\label{fig:ebv} Galactic dust extinction represented as
  $E(B-V)$ and plotted with degree resolution for Galactic latitudes
  $|b|>20^{\circ}$, together with the corresponding probability
  distribution function. This dust map peaks at $E(B-V)\sim 0.016$ mag,
  with a mean value of 0.048 mag and a median value of 0.028
  mag. Data taken from \citet{1998ApJ...500..525S}.}
\end{figure}

Astronomical data need to be corrected for extinction from dust in the
Milky Way.  Typical values (albeit with large variations) for the
reddening $E(B-V)\equiv A_B-A_V$ is $0.02$ with $A_V\sim 0.06$, see
Fig.~\ref{fig:ebv}. In this paper, we reinvestigate the relationship
between the Milky Way reddening and the dust maps generated in
\citet{1998ApJ...500..525S} using colours of quasars (QSOs) observed
within the Sloan Digital Sky Survey (SDSS). We also compare with
results from SDSS observations of brightest central galaxies (BCGs)
and luminous red galaxies (LRGs).  Compared with the original
investigation in \citet{1998ApJ...500..525S}, the improved statistics
and multiple band photometry give us the possibility to decrease the
errors on a linear relation while simultaneously constraining the
properties of Milky Way dust, in this case the total to selective
extinction ratio $R_V\equiv A_V/E(B-V)$.

We also performed a simple check to investigate if there are
deviations from the linear relationship, in other words whether
regions with high dust column density have more or less reddening than
the linear prediction. No such effect is found, nor do we find
any firm evidence of scatter in the dust properties as a function of
the line-of-sight.

While noting that there is still room for non-linear
behaviour, as well as scatter in dust properties at angular positions 
and scales not probed by the data, we conclude that at the positions and scales 
studied in this paper, it is a fair assumption that the reddening is linear
to the dust column density $D^T$ (in units of MJy/sr) with proportionality
constant $p=E(B-V)/D^T\sim 0.02$ and $R_V\sim 3$. The relative errors are
of the order of $10\,\%$ and are mainly limited by uncertainties in the filter response
functions and the source spectral templates.

This agrees well with the results of \citet{2011arXiv1111.4985B} where SDSS photometry of 
73 million stars is used to obtain the dust 
extinction towards each star giving a three-dimensional dust map of the 
Milky Way in the direction of the observed stripes. The results from this analysis are consistent 
with $R_V=3.0$, with a statistical and systematic error of $0.1$ and a scatter of $0.2$.

In \citet{2010ApJ...719..415P}, the colours of passively evolving galaxies from the 
SDSS are used to obtain corrections to the results of  
\citet{1998ApJ...500..525S}. They find that the reddening in 
regions of low dust temperature are underpredicted with maximum deviations in $E(B-V)$ of
45 millimagnitudes.

\citet{2010ApJ...725.1175S} use the blue edge of SDSS
main-sequence stars to recalibrate the reddening induced by the dust
maps of \citet{1998ApJ...500..525S}. They find that the reddening is
best described by a Fitzpatrick extinction curve with $R_V=3.1$,
whereas the extinction curve originally employed in
\citet{1998ApJ...500..525S} generically overestimates the reddening.
Similar results were obtained in \citet{2011ApJ...737..103S} where
very similar methods were used, but the reddening was measured by
comparing the observed colours of SDSS stars to the colours predicted
from their spectral features.

In the following, all limits are given as 95.4\,\% confidence levels (CL)
corresponding to $\chi^2 = \chi^2_{\rm min}+3.98$ for one parameter
and $\chi^2 = \chi^2_{\rm min}+6.16$ for two parameters.

\section{Method}\label{sec:method}
We used the temperature corrected dust maps from
\citet{1998ApJ...500..525S} to derive a reddening map by correlating
the dust maps to the observed colours of celestial objects, preferably
with a low intrinsic colour dispersion. The dust map, given as the
dust column density $D^T$ in units of MJy/sr, is derived from
  Infrared Astronomical Satellite (IRAS) data, in principle allowing
  for a resolution of approximately five arcminutes. However, these
  data are calibrated and temperature corrected using degree
  resolution Cosmic Background Explorer (COBE) Diffuse Infrared
  Background Experiment (DIRBE) data. We note that in
  \citet{2010ApJ...719..873K}, IRAS data are used to estimate the
  dust temperature in order to reach a resolution of five arcminutes
  also for the temperature corrected data of the Cygnus
  region. However, no all sky map data have yet been produced using
  this technique.

As a first attempt, we assumed that the observed colour,
e.g. $E(B-V)$, is proportional to the dust column density
\begin{equation}\label{eq:p}
  E(B-V)=p D^T + m,
\end{equation}
where we set out to derive the proportionality constant $p$, and $m$
represents an arbitrary offset that is marginalised over (possibly caused 
by calibration errors giving the wrong mean colour) and the dust column
density is given in units of MJy/sr, making $p$ dimensionless.
In \citet{1998ApJ...500..525S}, a value of the reddening parameter
$R_V=3.1$ was assumed, yielding a value of the proportionality constant
$p=0.0184 \pm 0.0014$ for Mg2-calibrated BCGs. 

Note that this can, in principle, be done for any set of
filters $X$ and $Y$
\begin{equation}\label{eq:kxy}
  E(X-Y)=k_{XY} D^T + m_{XY},
\end{equation}
and that multiple colour data thus can be used to infer dust properties.

\subsection{Fitting for $R_V$ and $p$}\label{sec:rvfit}
Since SDSS data allow for multiple colour combinations, we are also able to fit for the
properties of the Milky Way dust and relax any prior assumptions on the reddening parameter $R_V$. 
The SDSS $[u,g,r,i,z]$ filters have central wavelengths at $[3543,4770,6231,7625,9134]$
{\AA} and we thus expect $g-r$ and/or $u-r$ to correspond the most closely to
$B-V$. 

For each source and filter, we calculate the transmitted flux using a
redshifted source template. This is done for a grid of dust models in
order to compute how each dust model will affect a given colour. Note
that in doing this, we need to calculate the transmitted flux for each
dust model separately since $k$-corrections will depend on the dust
model. In practice, we have created tables that list the necessary
$k$-corrections over a grid of redshifts $A_V$ and $R_V$ for each
filter that we interpolate between. After calculating how a given
dust model affects each source template in each filter, we can compare
with the actual observed colours $E(X-Y)$ allowing us to constrain the
parameters $p$ and $R_V$.

\begin{figure}[!htb]
\resizebox{\hsize}{!}{\includegraphics{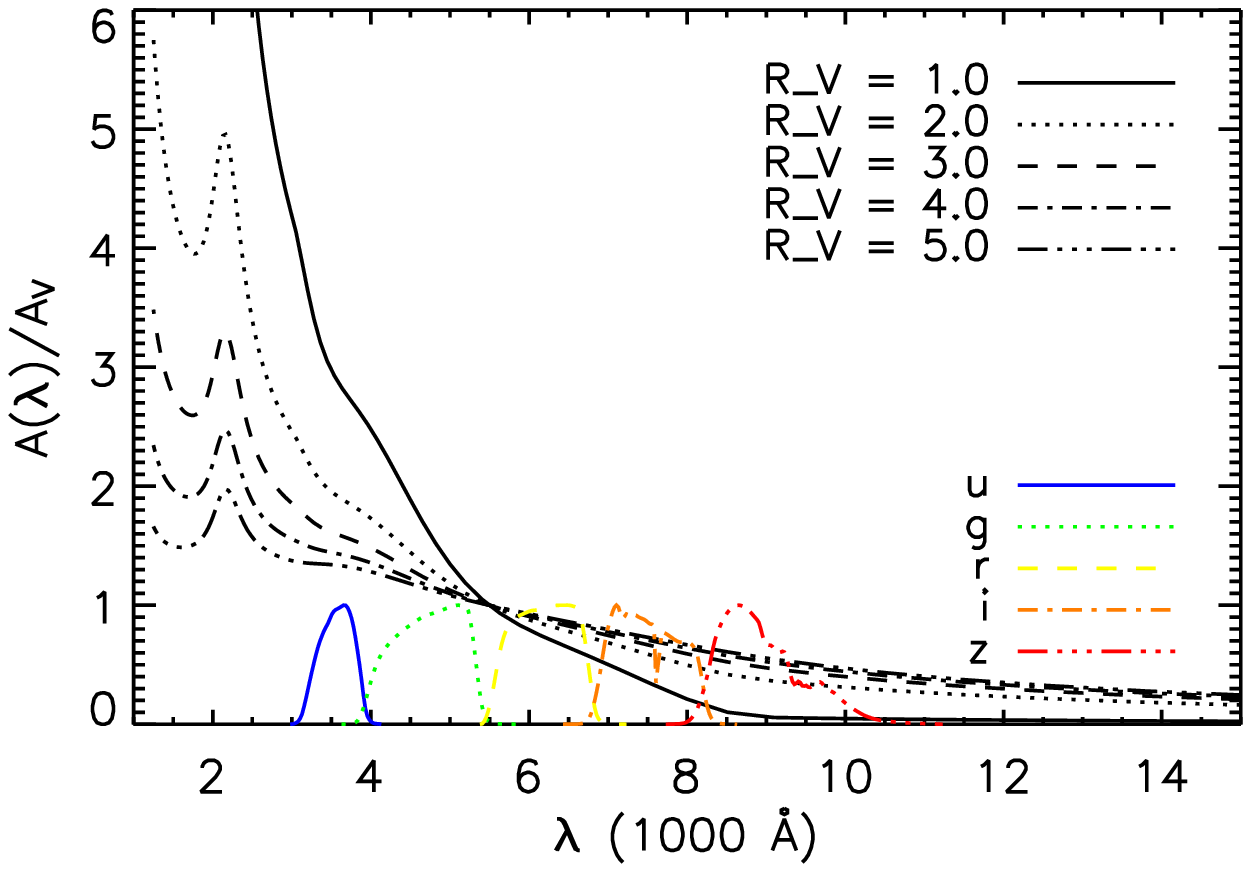}}
\resizebox{\hsize}{!}{\includegraphics{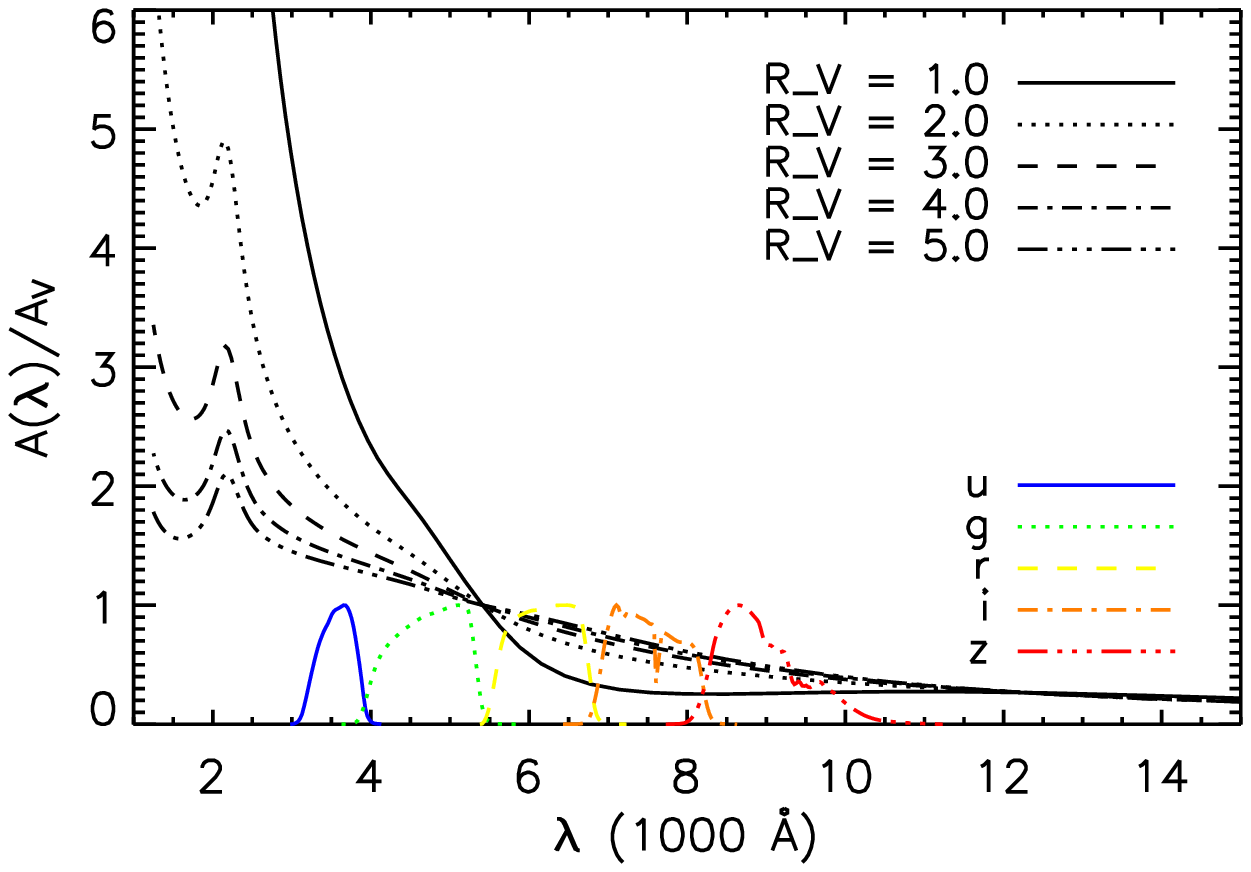}}
\caption{\label{fig:dustabs} Dust extinction curves for different values of $R_V$ shown together with the (arbitrary normalised) SDSS filter transmission curves. 
{\em Top:} Extinction curve for the CCM parametrisation.
{\em Bottom:} Extinction curve for the Fitzpatrick parametrisation.}
\end{figure}

Normalising such that zero column density corresponds to zero reddening, 
i.e. $m=0$ in Eq.~\ref{eq:p}, defining $f(\lambda)\equiv A(\lambda)/A_V$ and using
\begin{equation}
   A_V=pR_VD^T\, ,
\end{equation}
we can write
\begin{equation}
  \frac{A_1-A_2}{D^T}\equiv\frac{\Delta m}{D^T} = pR_V(f_1-f_2)\, ,
\end{equation}
and
\begin{equation}\label{eq:d}
  d\left(\frac{\Delta m}{D^T}\right ) = dpR_V(f_1-f_2)+pdR_V(f_1-f_2)+pR_Vd(f_1-f_2)\, ,
\end{equation}
showing how uncertainties in $D^T$, $\Delta m$, and the dust extinction
curve $f(\lambda)$ translates into uncertainties in $p$ and $R_V$.  In
the following, we will make use of the Milky Way extinction curves
$f(\lambda)$ as parametrised in \citet{1989ApJ...345..245C} (CCM) and
\citet{1999PASP..111...63F} (FTZ), plotted in Fig.~\ref{fig:dustabs}
together with the (arbitrarily normalised) SDSS filter transmission
curves. The extinction curve of \citet{1994ApJ...422..158O} gives
results very close to those derived using the CCM paramterisation.

\begin{figure}[!htb]
\resizebox{\hsize}{!}{\includegraphics{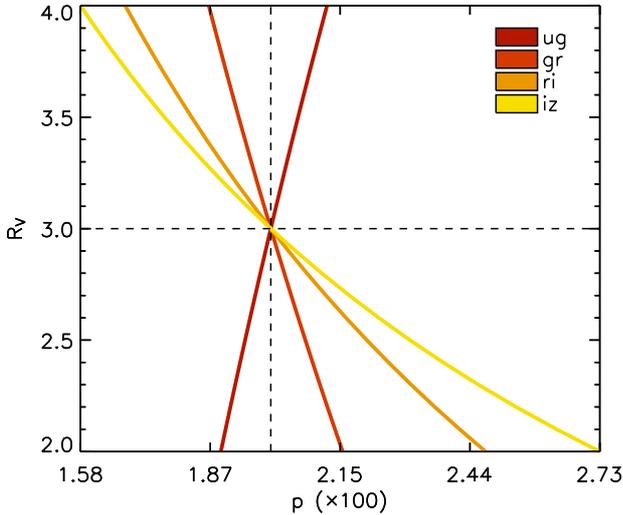}}
\caption{\label{fig:degen} Degeneracy between parameters $p$ and $R_V$ for SDSS colours $u-g$ (dark red), $g-r$ (red), $r-i$ (orange) and $i-z$ (yellow).
The degeneracy lines correspond to a reddening equivalent to $p=0.02$ and $R_V=3.0$. It is evident that multiple colours allow the degeneracy to be broken.}
\end{figure}

We now turn to the expected degeneracy between $R_V$ and $p$ and how
this can be broken using multiple colour data.  In the CCM-model, we
have $f=a+b/R_V$ and $f_1-f_2=(a_1-a_2)+(b_1-b_2)/R_V\equiv \Delta
a+\Delta b/R_V$. Setting $d\left(\Delta m/D^T\right )=0$ in
Eq.~\ref{eq:d} and putting $d(\Delta a)=d(\Delta b)=0$, i.e. no
systematic error in the extinction curves, we obtain
\begin{equation}
  \frac{dp}{p} = \frac{-dR_V}{R_V+\Delta b/\Delta a}\, .
\end{equation}
This can be integrated to give
\begin{equation}
  p = p_0\frac{R_{V,0}+\Delta b/\Delta a}{R_{V}+\Delta b/\Delta a}\, ,
\end{equation}
describing the relation between different values of $R_V$ and $p$
giving the same extinction per unit column density $\Delta m/D^T$.
The degeneracy lines for the SDSS standard colours $u-g$ (dark red),
$g-r$ (red), $r-i$ (orange) and $i-z$ (yellow) are plotted in
Fig.~\ref{fig:degen}, displaying the power of multiple colour data to
break the degeneracy between $p$ and $R_V$. We see that $p$ is
anti-correlated with $R_V$, if the difference $(A_X-A_Y)/A_V$ is
decreasing more slowly than $1/R_V$ with $R_V$, which is the case for the
three reddest SDSS standard colours.  The slopes of these curves
compare well with the confidence contours derived from data in the
forthcoming sections.


\section{Error analysis}\label{sec:syseff}
We first note that constant filter calibration errors do not affect our
results, since we are marginalising over any constant offset between
the observed and theoretically calculated colours.  However, we may
get systematic effects if the spectral template is wrong and/or if the
shape of the filter response function is incorrectly modeled since this
will change the effective wavelength of the filters. We assume that
this will induce fractional uncertainties in the effective wavelengths
of [0.02, 0.01, 0.01, 0.01, 0.01] in $[u,g,r,i,z]$ filters
\citep{2010AJ....139.1628D}.  We note that this error will be a true
systematic error since the effect will be correlated between all source
redshifts and colours.  For comparison, for our QSO template,
effective wavelengths will vary by $\sim 2-3$\,\% when varying the
source redshift and dust parameters. \citet{1998ApJ...500..525S}
calculate their reddening estimates to have a dispersion of $\sim
16\,\%$. Attributing a major part of this dispersion to a
scatter in the dust column density, we assume a fractional uncertainty
of 10\,\% in the dust map values $D^T$. Since this uncertainty is
treated as a statistical error, and we are systematics dominated, the
exact value of this fractional error does not affect the end
result.
A possible problem not taken into account in 
this paper is that the dust map may be contaminated by the
extragalactic background at low values \citep{2007PASJ...59..205Y}.

We incorporate uncertainties using a Monte Carlo-based approach as follows:
\begin{itemize}
  \item We pick a random value for the shift in the effective wavelength for each 
        filter and compute the corresponding colour offset. We also randomise 
        the value for $D^T$ at each sky position within the given uncertainties.  
  \item We derive the likelihood over a grid of fitted parameters. In doing this,
        we use the statistical sources of error from observational uncertainties and
        the intrinsic colour dispersion of the sources.
  \item We repeat $N$ times (we use $N\sim 50$) to obtain $N$ likelihood functions.	 
  \item We compute the mean of the $N$ likelihood functions.	 
\end{itemize}
In this way we take full account of the different correlations between sources of error, and the resulting 
confidence contours correctly incorporate both statistical and systematic effects. 
As noted in Sect.~\ref{sec:introduction}, the dominating source of error is the possibility of shifts in the effective filter wavelengths,
originating in uncertainties in the filter response
functions and the source spectral templates.

\section{Quasar analysis}\label{sec:qso}
\begin{figure}[!htb]
\resizebox{\hsize}{!}{\includegraphics{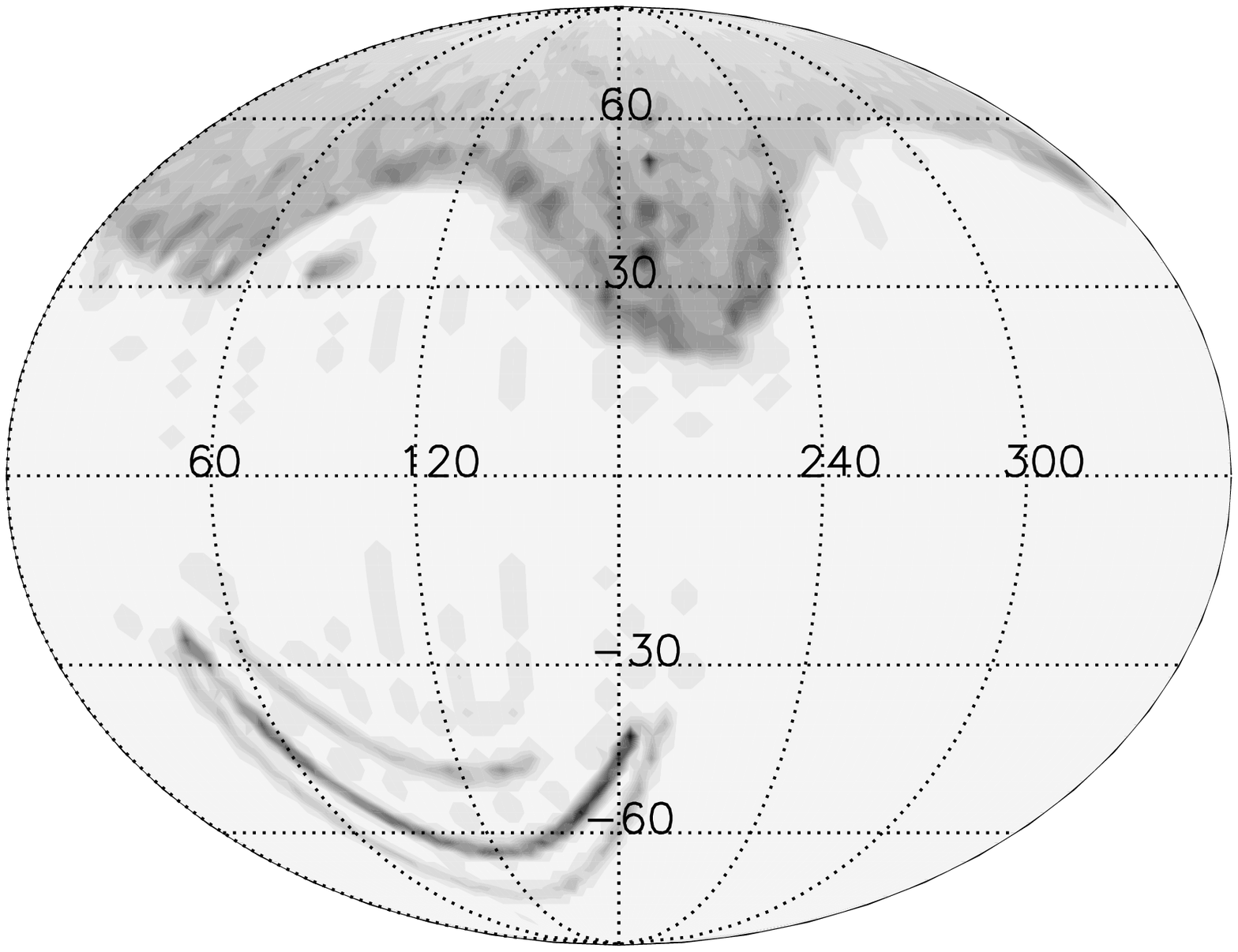}}
\resizebox{\hsize}{!}{\includegraphics{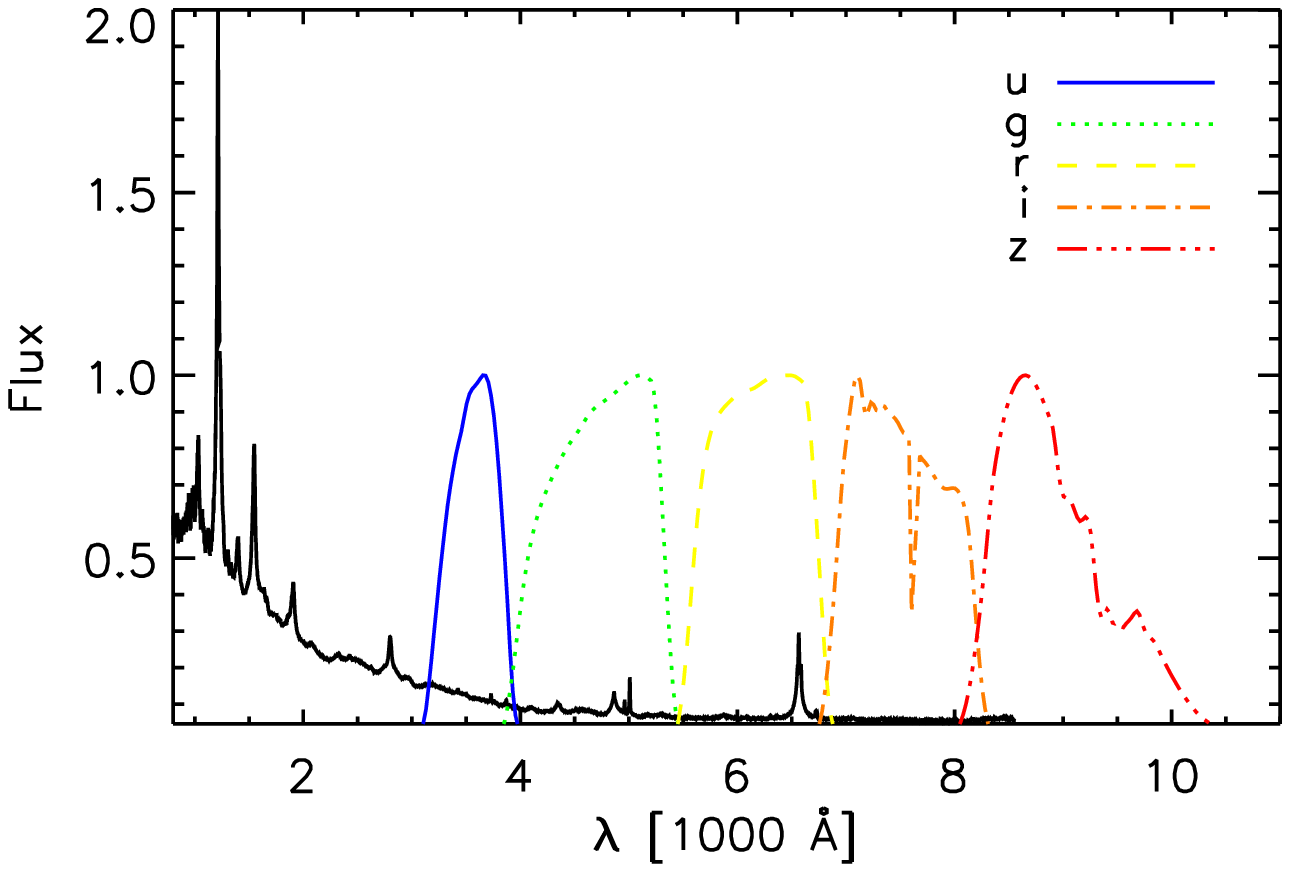}}
\caption{\label{fig:qsoposspec} {\em Top:} Angular positions of the SDSS QSOs used in this analysis with darker colours corresponding to 
higher observed surface densities of QSOs, reflecting the given survey strategy. 
{\em Bottom:} Restframe QSO template spectrum shown with arbitrary normalised SDSS filter transmission curves.}
\end{figure}

It is well known that QSOs form a relatively homogeneous group of
objects in terms of colours and spectral features. They also have
large flux at short wavelengths which means that we can use blue
colours, e.g. $u-r$ even for high redshift sources, which allows us to
effectively break the degeneracy between $p$ and $R_V$ as shown in
Fig.~\ref{fig:degen}.  We use the fifth edition of the Sloan Digital
Sky Survey Quasar Catalog \citep{2007AJ....134..102S}
consisting of 77\,429 objects.  The area covered by the catalogue is
5\,740 deg$^2$, see Fig.~\ref{fig:qsoposspec}.
QSOs have also been used successfully to constrain the amount and properties of dust in galaxies at 
cosmological distances \citep{2006A&A...450..971O,2008A&A...485..403O} and for intergalactic dust 
\citep{2005JCAP...02..005O}.

We are primarily interested in the colours of QSOs, so it is 
important to understand how the objects in the SDSS QSO
catalogue are chosen in order to assess whether there are
selection effects that need to be considered.
The QSO candidate selection is
based on the position of sources in multidimensional SDSS colour
space. In terms of colours, QSOs are mainly identified as being outside 
the four-dimensional stellar loci points in $[u,g,r,i,z]$ space 
\citep{2002AJ....123.2945R}. Since the method is purely
differential, any general colour bias will not affect the end
result; however, we need to make sure that the selection does not
depend on the dust column. Since the selection is made after
correcting for the extinction maps in \citet{1998ApJ...500..525S},
assuming that these provide a fair model for the extinction, there is
no reason that the selection should be severely affected by Milky Way
dust extinction. By applying an extra colour cut of 
included objects where the (uncorrected) colour is less than
$2\,\sigma$ off the mean value, as evaluated in bins of different dust
column density, we are guaranteed a sample not correlated with
the dust column, unless initial dust corrections are grossly
incorrect.

Only QSOs in the redshift range $0.5<z<3.0$ are considered as QSOs outside this range
are less accurately described by the spectral
template used, created by combining the Hubble Space Telescope radio-quiet composite spectrum
\citep{2002ApJ...565..773T} with the SDSS median composite
spectrum \citep{2001AJ....122..549V} as described in \citet{2006A&A...450..971O,2008A&A...485..403O}. 
Generally, all colours are relatively homogeneous at $z\la 2$. Typically, at $z\sim 1$, we have
$\sigma (u-g)=0.15$, $\sigma (g-r)=0.13$, $\sigma (r-i)=0.093$, and
$\sigma (i-z)=0.11$.
However, when the Lyman-$\alpha$ emission is being redshifted into the 
observed filters, the dispersion increases. We thus cut the $u$-band data at
$z>3000/1210-1\sim 1.5$ and the $g$-band data at $z>4000/1210-1\sim 2.3$.

\begin{figure*}[!htb]
\centering
\includegraphics[width=17cm]{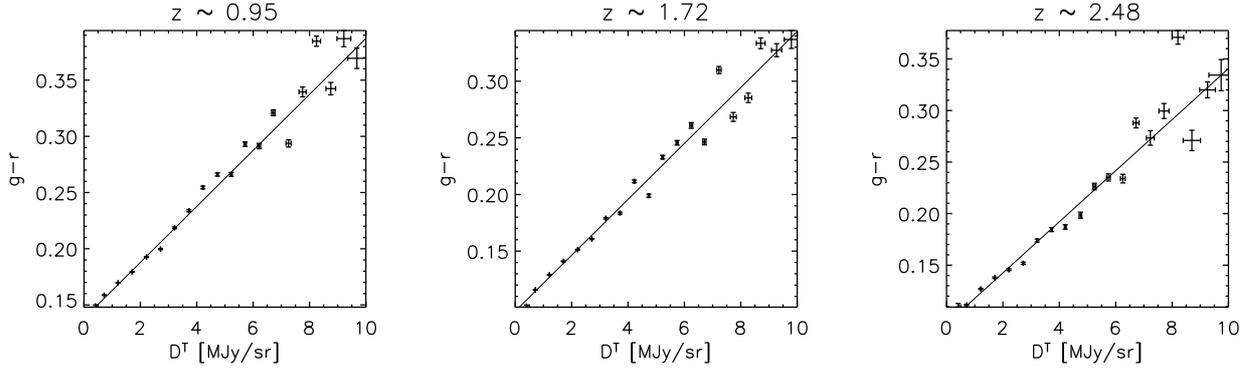}
\caption{\label{fig:kbcg} Observed $g-r$ colour as a function of Galactic dust column density for SDSS QSOs at three different redshifts. 
Note the almost perfect linear relationship between the observed colour and the dust density.}
\end{figure*}

\begin{figure}[!htb]
\resizebox{\hsize}{!}{\includegraphics{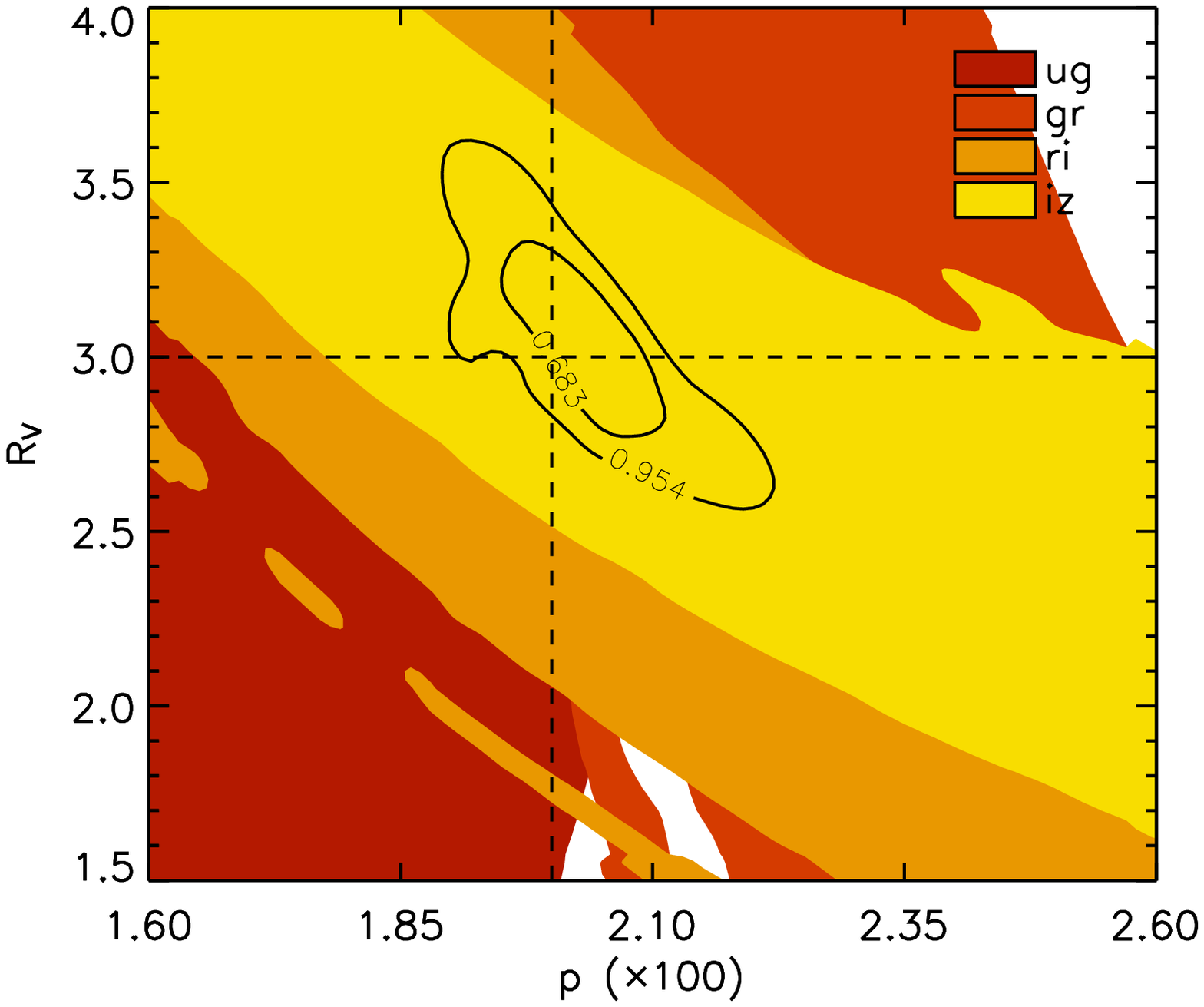}}
\resizebox{\hsize}{!}{\includegraphics{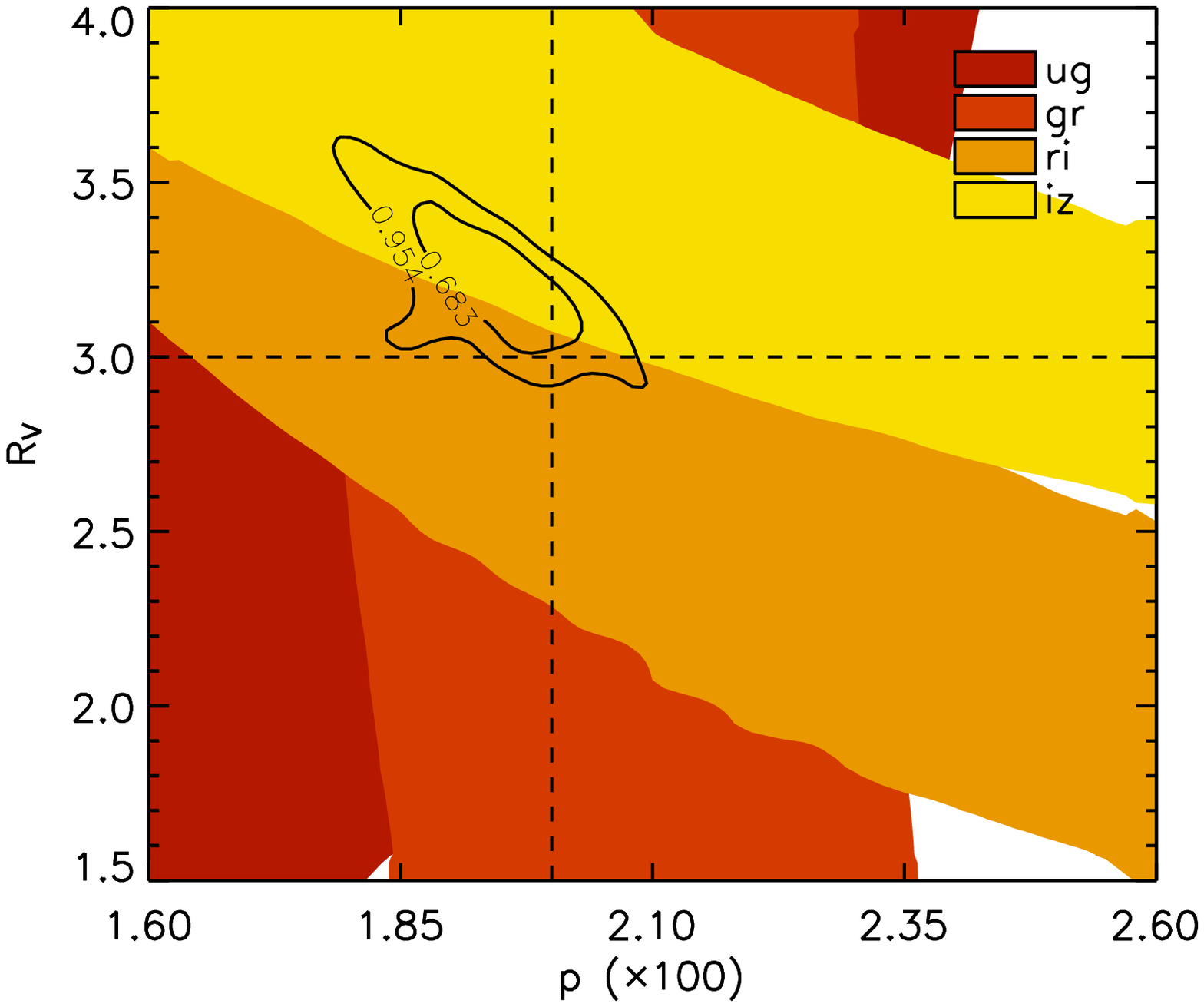}}
\caption{\label{fig:qsorv} Constraints on $p$ and $R_V$ as derived from the observed colours of SDSS QSOs and the measured dust column density at the angular position
of each QSO. Solid colours correspond to 95.4\,\% CL for two parameters. {\em Top:} Results for CCM dust corresponding to 95.4\,\% CL for one parameter of $p=0.0205 \pm 0.0015$ and $R_V=3.10 \pm 0.43$. 
{\em Bottom:} Results for Fitzpatrick dust corresponding to 95.4\,\% CL for one parameter of
$p=0.0190 \pm 0.0015$ and $R_V=3.25 \pm 0.37$.}
\end{figure}

In order to display the close to linear relationship between the observed QSO colour and the dust density, 
in Fig.~\ref{fig:kbcg} we show the observed QSO $g-r$ colour as a function of 
galactic dust column density at redshifts $z\sim 0.95$, $z\sim 1.72$ and $z\sim 2.48$. 
Results for the full fit of $p$ and $R_V$
are shown in Fig.~\ref{fig:qsorv} for the CCM dust extinction parametrisation (top panel) and the 
Fitzpatrick parametrisation (bottom panel). Solid colours correspond  to 95.4\,\% CL for two parameters. 
Our constraints are $p=0.0205 \pm 0.0015$ and $R_V=3.10 \pm 0.43$ (CCM)
and $p=0.0190 \pm 0.0015$ and $R_V=3.25 \pm 0.37$ (FTZ)
(95.4\,\% CL for one parameter).
Although consistent at the 95.4\,\% CL,
compared to the $r-i$ and $i-z$ colours, we have on average less extinction in $u-g$
and more extinction in $g-r$ for the CCM parametrisation. For
Fitzpatrick dust, compared to $g-r$ and $r-i$, $u-g$ has less
extinction and $i-z$ more extinction on average.

\begin{figure}[!htb]
\resizebox{\hsize}{!}{\includegraphics{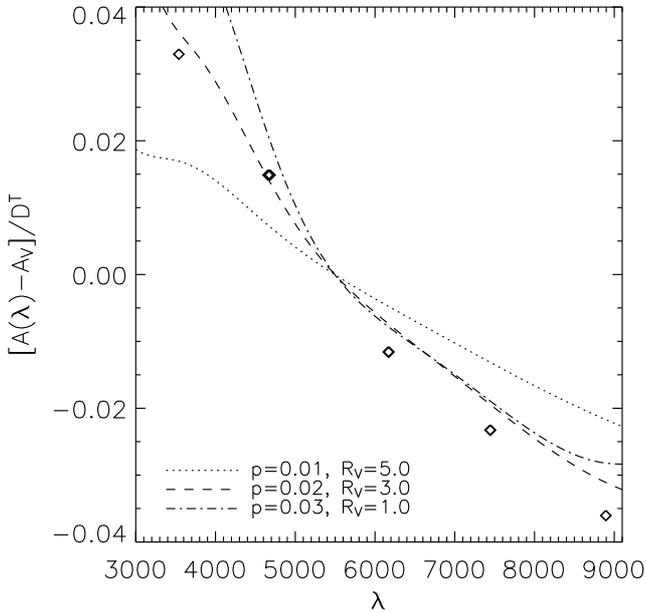}}
\caption{\label{fig:qsod3} Empirical dust law as derived from QSO colours (diamonds) compared to the extinction curves of CCM dust with $p=0.01, R_V=5.0$ (dotted curve), 
$p=0.02, R_V=3.0$ (dashed curve), and $p=0.03, R_V=1.0$ (dash-dotted curve).}
\end{figure}

Albeit with very limited resolution, multiple colour data allow us to
derive the dust extinction curve directly from the data, without the
need to assume any specific dust extinction parametrisation. For each
colour, we fit the slope $k_{XY}$ in Eq.~\ref{eq:kxy}, apply the
necessary $k$-corrections, and normalise the extinction to zero in
the $V$-band.  In Fig.~\ref{fig:qsod3} we show the dust law as
derived from QSO colours compared to the extinction curves for some
fiducial dust models. Assuming the errors to be accurately represented
by the uncertainty of the slope in the linear fit, the error bars are
too small to be visible. However, this assumption is valid only in the
case that the relation between the reddening and the dust column
density indeed is linear and constant over the field. Furthermore, the
dominating source of error of possible systematic shifts in the
effective central filter wavelengths have not been included. Note that
since the normalisation is arbitrary, only the shape of the derived
extinction curve is relevant.

\section{Brightest central galaxies analysis}\label{sec:sdssbcg}
The brightest galaxies at the centre of galaxy clusters are
empirically known to be relatively homogeneous in colour.  Since
high-luminosity early-type galaxies are generally redder than
low-luminosity galaxies (usually attributable to higher metallicity),
the colour dispersion can be further reduced using colour-magnitude
relations. In this paper, however, we defer from employing any such
calibration.

\begin{figure}[!htb]
\resizebox{\hsize}{!}{\includegraphics{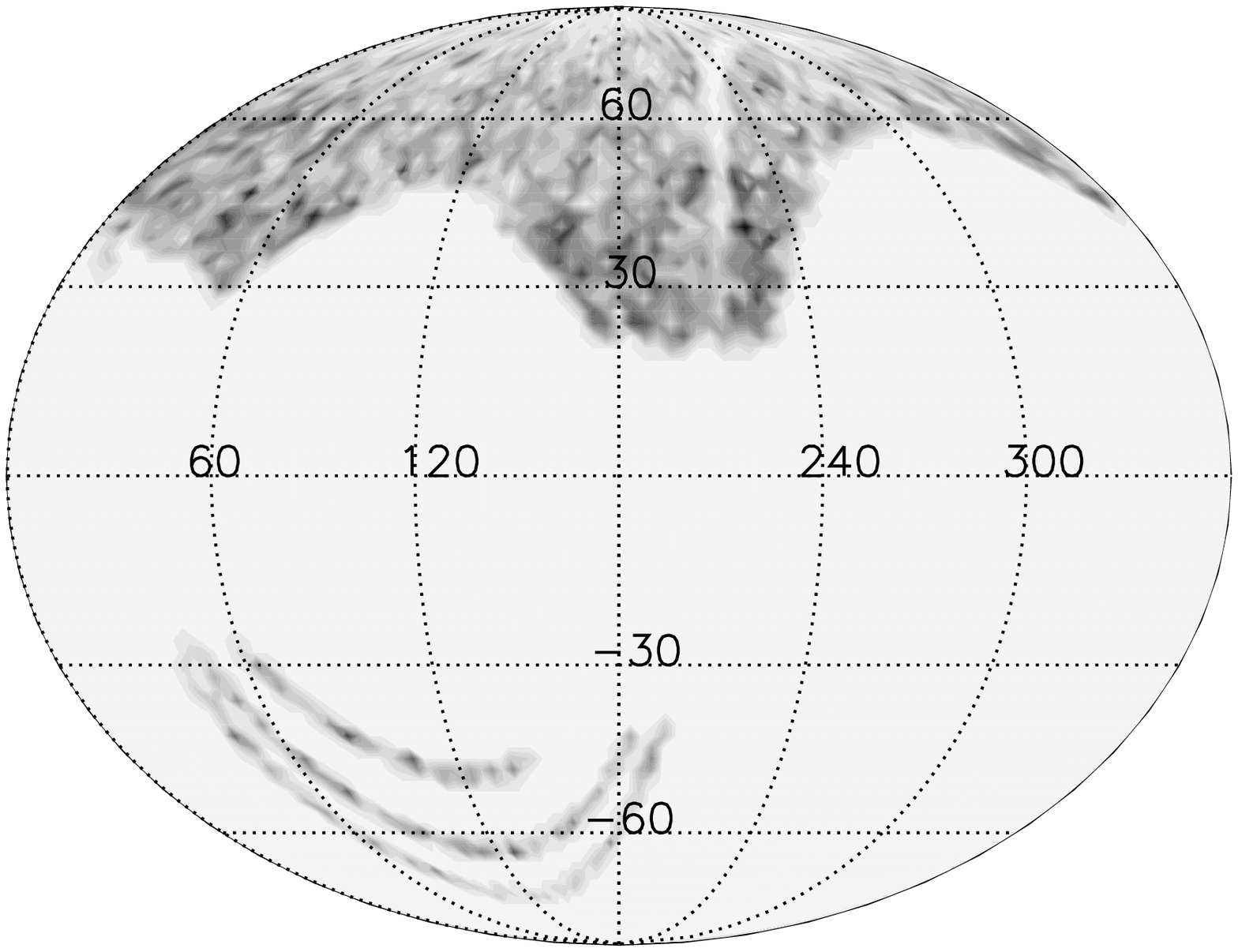}}
\resizebox{\hsize}{!}{\includegraphics{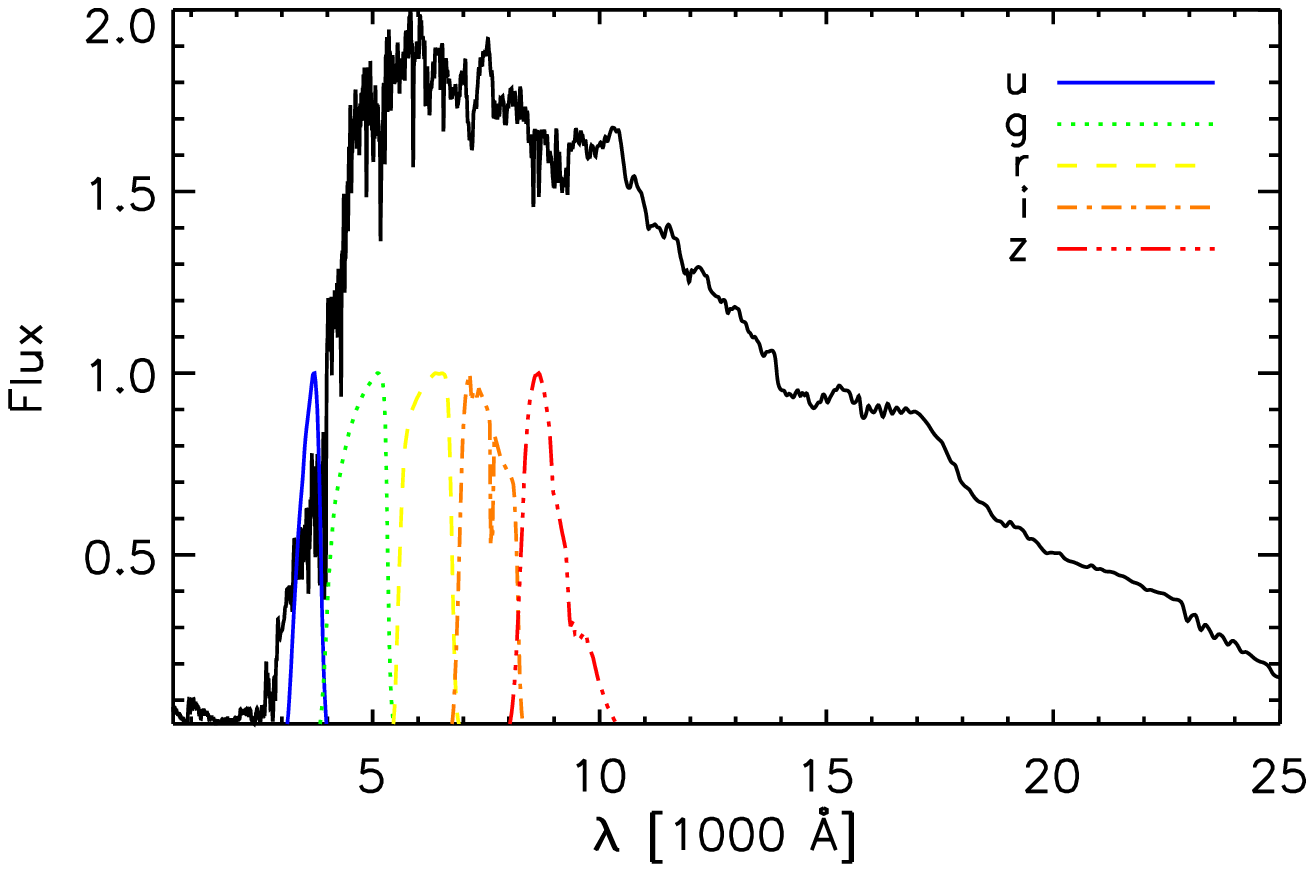}}
\caption{\label{fig:bcgposspec} {\em Top:} Angular positions of the SDSS BCGs used in this analysis with darker colours corresponding to 
higher observed surface densities of BCGs.
{\em Bottom:} Rest-frame BCG template spectrum shown together with arbitrarily normalised SDSS filter transmission curves. 
Note that the same template is used for the LRG analysis. }
\end{figure}

We have used data from the MaxBCG catalogue of 13\,823 galaxy
clusters from the Sloan Digital Sky Survey
\citep{2007ApJ...660..239K}. In order to extract the full photometric
data for the brightest central galaxies (BCGs), we retrieved all objects
within 6 arcseconds from the BCG positions. We managed to identify and
retrieve $[u,g,r,i,z]$ data for 13\,796 BCGs with redshifts
$0.046<z<0.37$. The angular positions of the BCGs are depicted in
Fig.~\ref{fig:bcgposspec}. Generally, all colours involving the $u$-band
are not very well-determined. Typically (at $z\sim 0.2$), we have
$\sigma (u-g)=0.32$, $\sigma (g-r)=0.059$, $\sigma (r-i)=0.029$, and
$\sigma (i-z)=0.042$.

\begin{figure}[!htb]
\resizebox{\hsize}{!}{\includegraphics{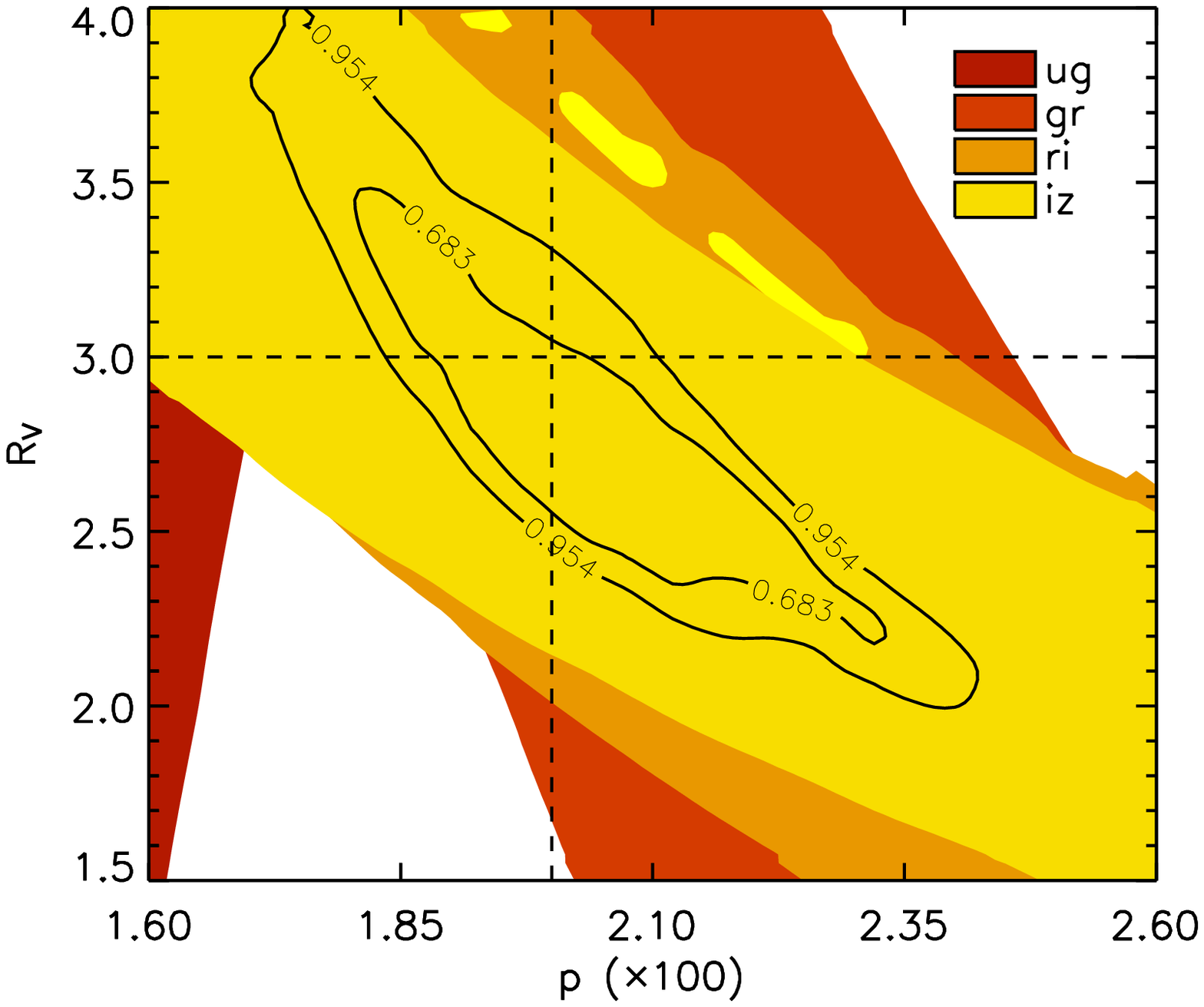}}
\resizebox{\hsize}{!}{\includegraphics{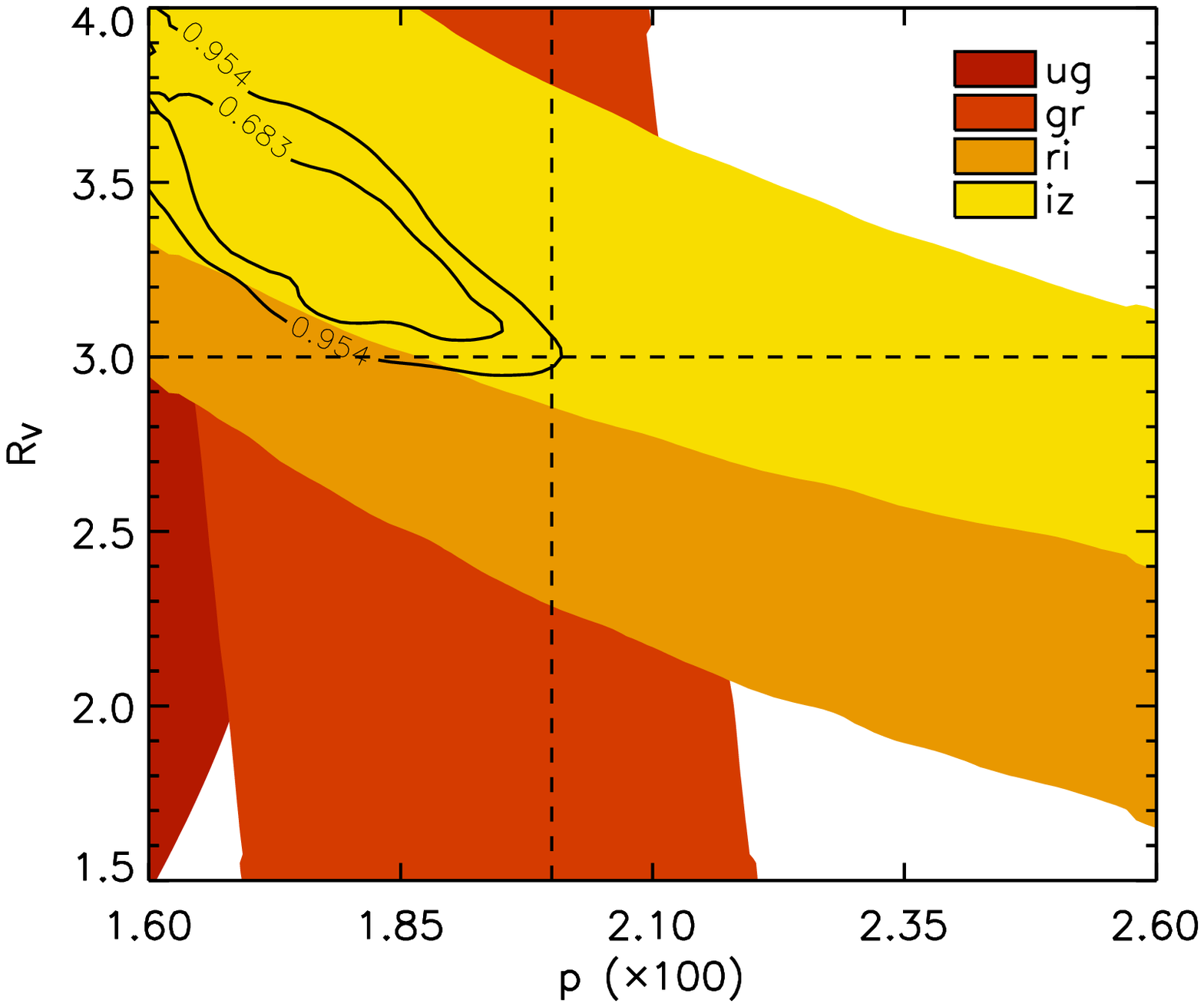}}
\caption{\label{fig:bcgrv} Constraints on $p$ and $R_V$ as derived from the observed colours of SDSS BCGs and the measured dust column density at the angular position
of each BCG. Solid colours correspond to 95.4\,\% CL for two parameters. {\em Top:} Results for CCM dust corresponding to 95.4\,\% CL for one parameter of $p=0.0205 \pm 0.0030$ and $R_V=2.89 \pm 0.88$. 
{\em Bottom:} Results for Fitzpatrick dust corresponding to 95.4\,\% CL for one parameter of
$p=0.0177 \pm 0.0020$ and $R_V=3.42 \pm 0.48$.}
\end{figure}

\begin{figure}[!htb]
\resizebox{\hsize}{!}{\includegraphics{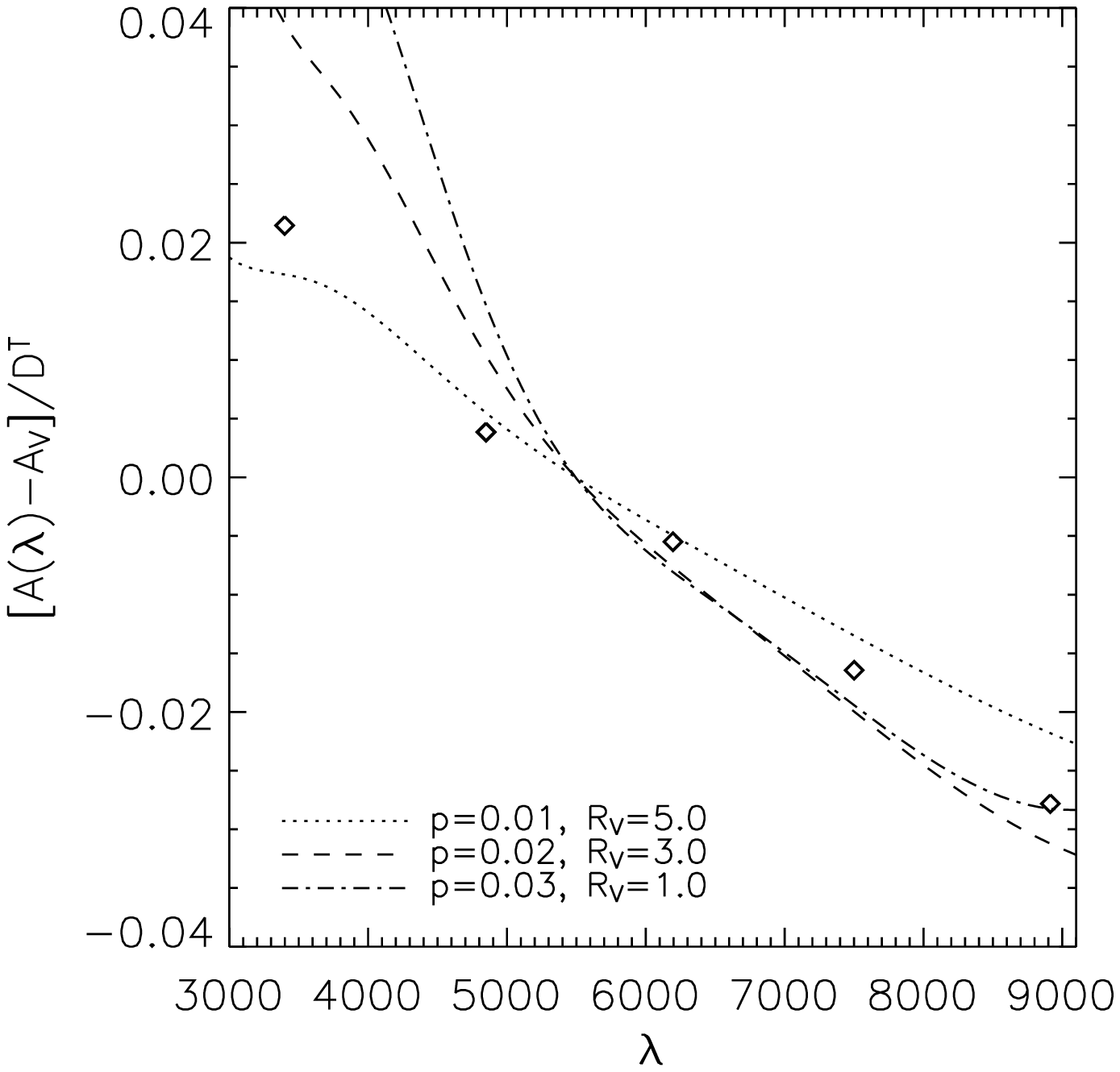}}
\caption{\label{fig:bcgd3} Empirical dust law as derived from BCG colours (diamonds) compared to the extinction curves of CCM dust with $p=0.01, R_V=5.0$ (dotted curve), 
$p=0.02, R_V=3.0$ (dashed curve), and $p=0.03, R_V=1.0$ (dash-dotted curve).}
\end{figure}

We cut all colours involving the the $u$-band at $z>0.15$ since the
colour dispersion becomes very large at those redshifts.  For
consistency with the QSO and LRG analysis, we reject objects with
uncorrected colours more than $2\,\sigma$ off the mean value, as
evaluated in bins of different dust column density. Results for
fitting $p$ and $R_V$ to the BCG data are shown in
Fig.~\ref{fig:bcgrv}.  Our constraints are $p=0.0205 \pm 0.0030$ and
$R_V=2.89 \pm 0.88$ (CCM), and $p=0.0177 \pm 0.0020$ and $R_V=3.42 \pm
0.48$ (FTZ) (95.4\,\% CL for one parameter). Like the QSO
constraints, compared to $r-i$ and $i-z$ that give $p=0.02$ for
$R_V=3$, we have less extinction in $u-g$ and more extinction in $g-r$
for the CCM parametrisation. For Fitzpatrick dust, $g-r$ and $r-i$
give $p=0.02$ at $R_V=3$ while $u-g$ has less extinction and $i-z$
more extinction. In Fig.~\ref{fig:bcgd3} we show the empirically
derived dust law.

\section{Luminous red galaxies}\label{sec:lrg}
\begin{figure}[!htb]
\resizebox{\hsize}{!}{\includegraphics{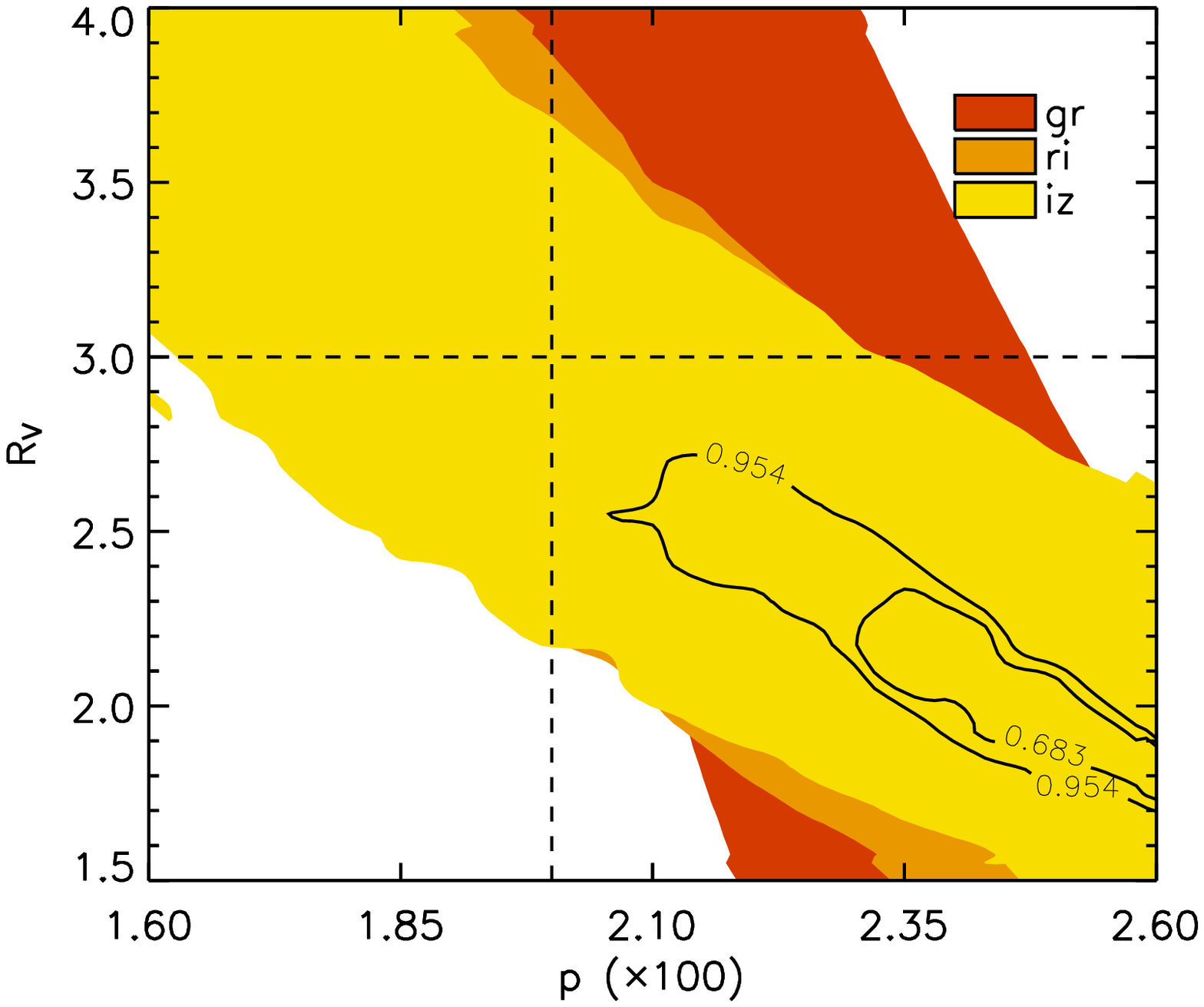}}
\resizebox{\hsize}{!}{\includegraphics{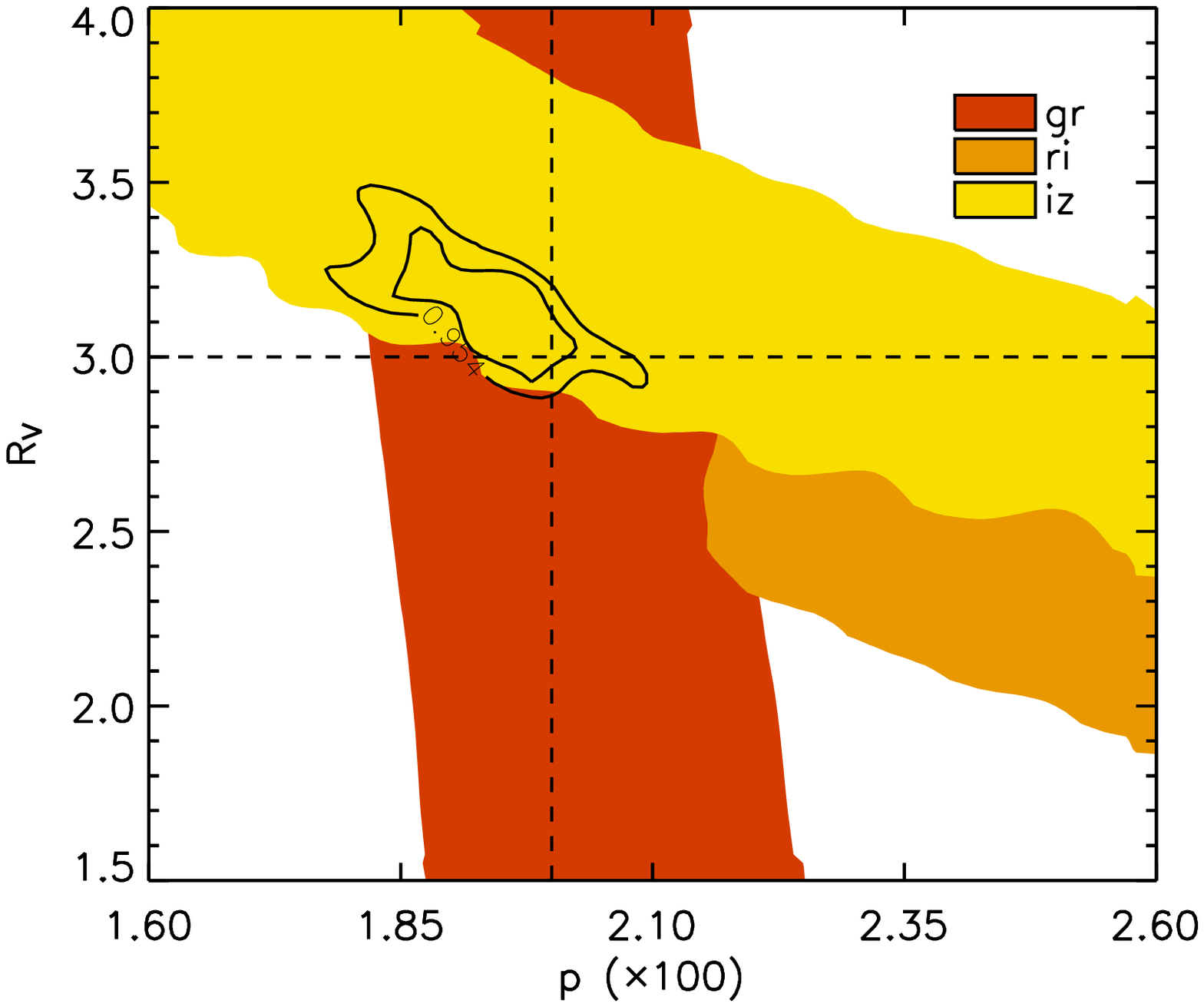}}
\caption{\label{fig:lrgrv} Constraints on $p$ and $R_V$ as derived from the observed colours of SDSS LRGs and the measured dust column density at the angular position
of each LRG. Solid colours correspond to 95.4\,\% CL for two parameters.{\em Top:} Results for CCM dust corresponding to 95.4\,\% CL for one parameter of $p=0.026 \pm 0.004$ and $R_V=1.95 \pm 0.35$. 
{\em Bottom:} Results for Fitzpatrick dust corresponding to 95.4\,\% CL for one parameter of
$p=0.019 \pm 0.0025$ and $R_V=3.16 \pm 0.36$.}
\end{figure}

\begin{figure}[!htb]
\resizebox{\hsize}{!}{\includegraphics{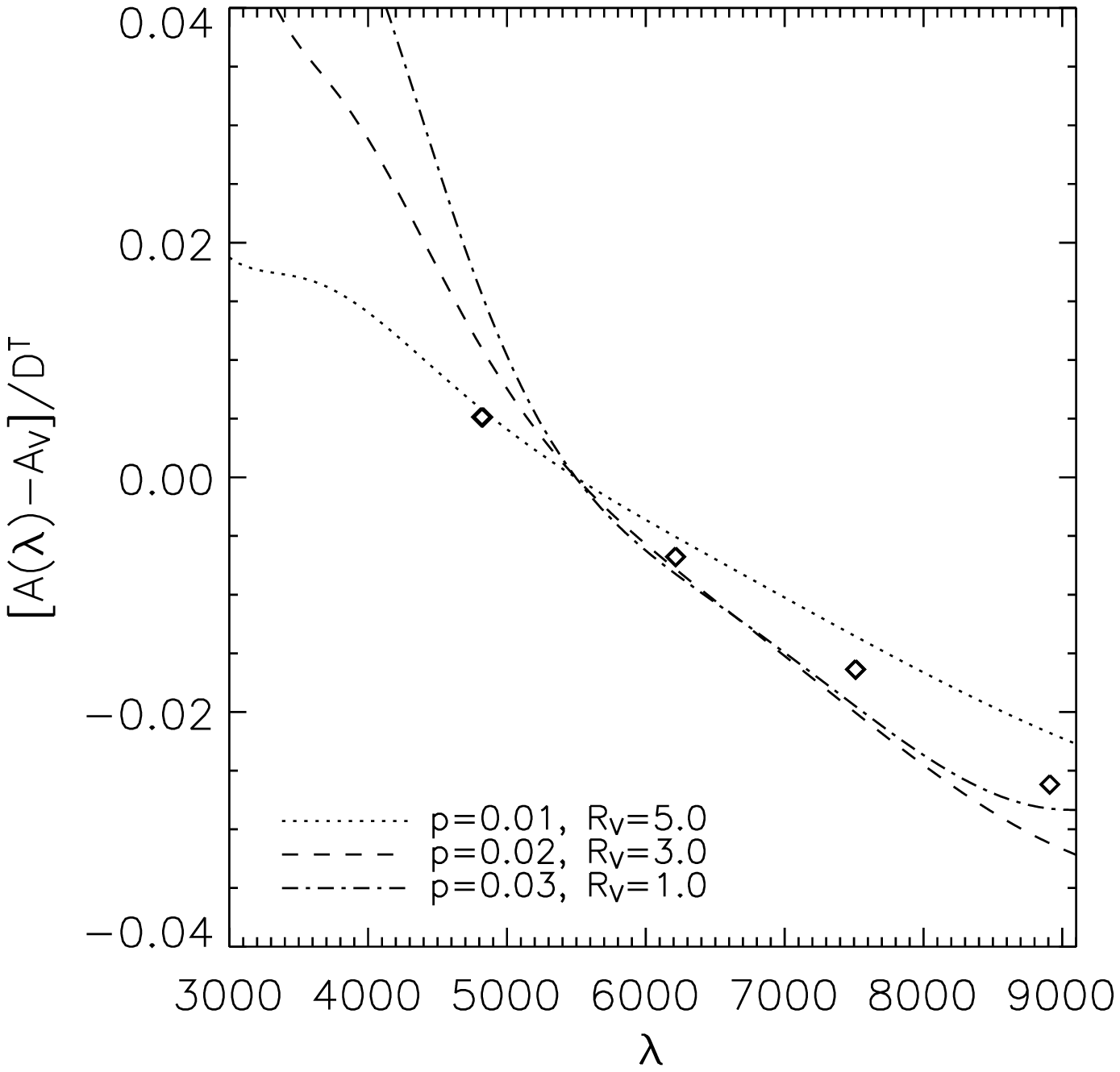}}
\caption{\label{fig:lrgd3} Empirical dust law as derived from LRG colours (diamonds) compared to the extinction curves of CCM dust with $p=0.01, R_V=5.0$ (dotted curve), 
$p=0.02, R_V=3.0$ (dashed curve), and $p=0.03, R_V=1.0$ (dash-dotted curve).}
\end{figure}

Since luminous red galaxies (LRGs) are also known to be
homogeneous in terms of colours, we have extracted data for 114\,151
LRGs from the SDSS data base. Generally, all colours involving the
$u$-band are not well-determined. Typically, at $z\sim 0.3$ we have
$\sigma (u-g)=0.84$, $\sigma (g-r)=0.086$, $\sigma (r-i)=0.043$, and
$\sigma (i-z)=0.056$.  We therefore do not use any colours involving
the $u$-band in this analysis. As discussed in
  Sect.~\ref{sec:qso}, we reject objects with uncorrected colours more
  than $2\,\sigma$ off the mean value, as evaluated at different dust
  column values, in order for the final sample to be unbiased by dust
  extinction. Fitting for $p$ and $R_V$ gives the results shown in
Fig.~\ref{fig:lrgrv}.  Our constraints are $p=0.026 \pm 0.004$ and
$R_V=1.95 \pm 0.35$ (CCM), and $p=0.019 \pm 0.0025$ and $R_V=3.16 \pm
0.36$ (FTZ) (95.4\,\% CL for one parameter). Compared to the QSO and
BCG analysis, the absence of the $u$-band pushes best-fit values
towards higher $p$ and lower $R_V$.  Assuming $R_V=3$, compared to
$r-i$ and $i-z$ that give $p=0.02$, we have more extinction in $g-r$
for the CCM parametrisation. For Fitzpatrick dust, $r-i$ gives
$p=0.02$ while $g-r$ and $i-z$ have more extinction.  In
Fig.~\ref{fig:lrgd3} we show the derived dust law.

\section{Discussion}\label{sec:disc}
\begin{figure}[!htb]
\resizebox{\hsize}{!}{\includegraphics{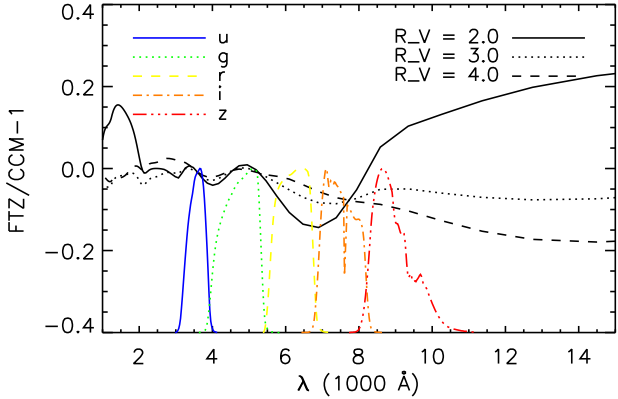}}
\caption{\label{fig:dustcomp} The
differences between the CCM and FTZ dust extinction curves.}
\end{figure}

As seen in the previous sections, there are slight differences in the preferred values 
of $p$ and $R_V$, depending on the dust parametrisation and the data set used. 
Generally, for the CCM parametrisation with $R_V=3$, for $r-i$ and
$i-z$ we get $p=0.02$.  We have more extinction (higher $p$) in $g-r$
and less in $u-g$. For Fitzpatrick dust, $r-i$ gives $p=0.02$ while
$g-r$ and $i-z$ have more extinction and $u-g$ less. How can we
understand these differences?  First, we need to understand the
differences between the CCM and FTZ dust extinction parametrisations. 
In Fig.~\ref{fig:dustcomp}, we plot
\begin{equation}
  \frac{f(\lambda)_{FTZ}-f(\lambda)_{CCM}}{f(\lambda)_{CCM}}\, .
\end{equation}
The reason that the we need higher $p$ for $i-z$ for the FTZ model is
that the extinction curve is flatter there compared to the CCM
model. We also see that it is steeper for $g-r$ which means that we
get a lower $p$ for that colour. That is, we need a dust law that is
slightly steeper for $g-r$ and flatter for $u-g$ compared to CCM. As
for the differences between the data sets, comparing BCGs and LRGs, we
see that the differences can be explained by having a flatter template
spectrum for LRGs in the region covering $g-r$, compared to the BCG
template.  Comparing BCGs and QSOs, the differences can be resolved by
having either the BCG template flatter for $u-g$ and $g-r$ or the QSO
template steeper in the same wavelength region.

\section{$R_V$-variations}\label{sec:rv-var}
\begin{figure*}[!htb]
\centering
\includegraphics[width=8cm]{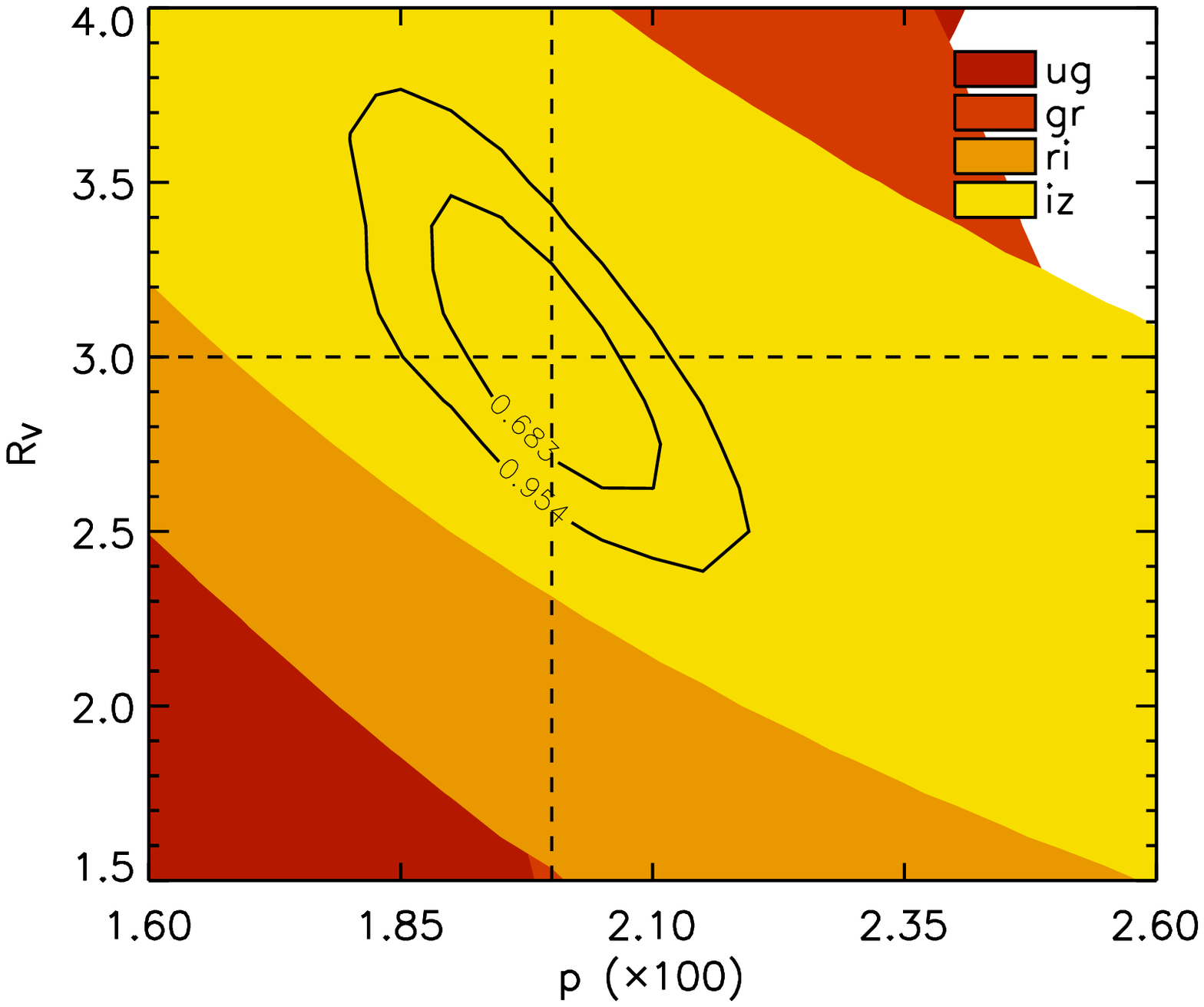}
\includegraphics[width=8cm]{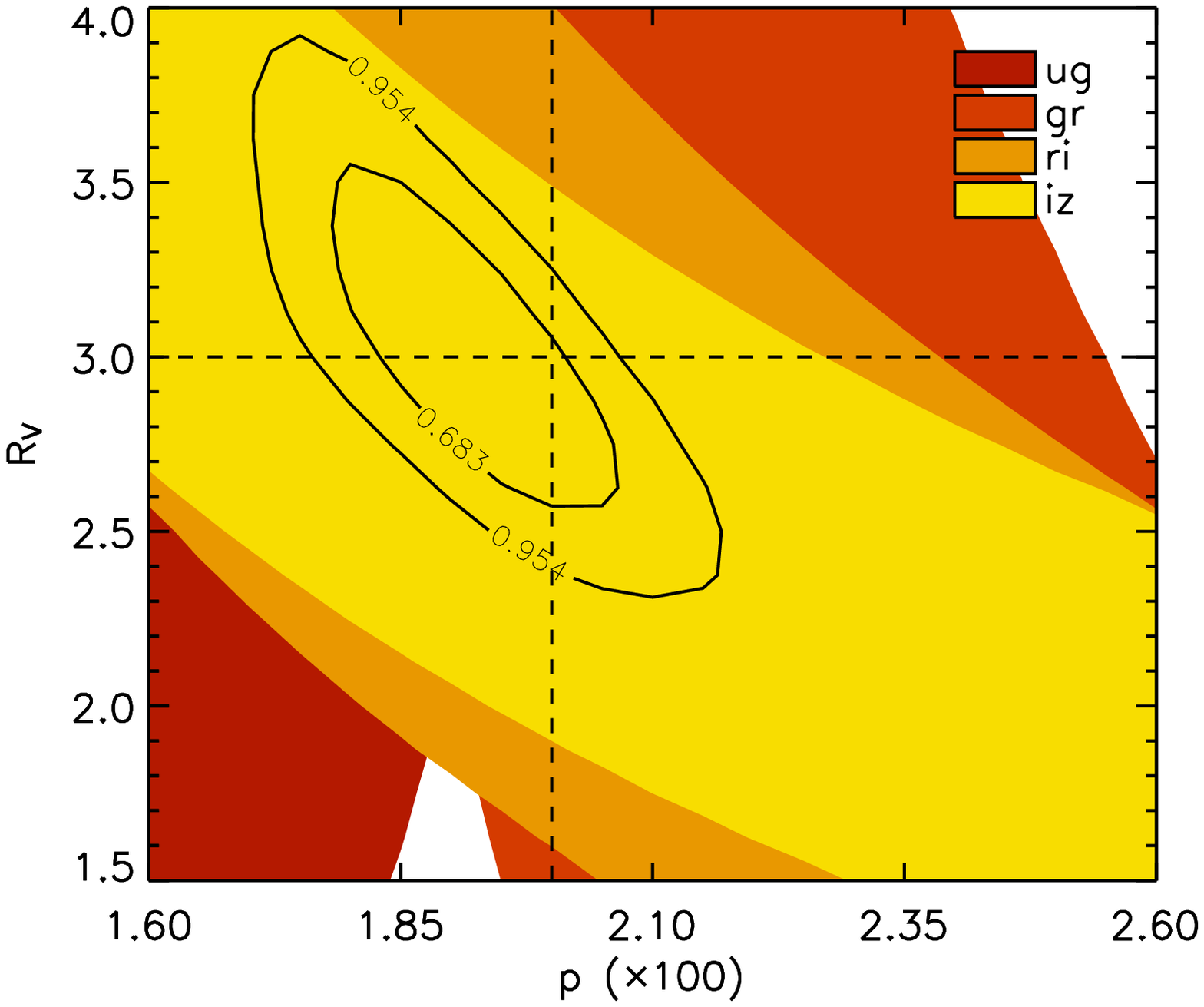}
\includegraphics[width=8cm]{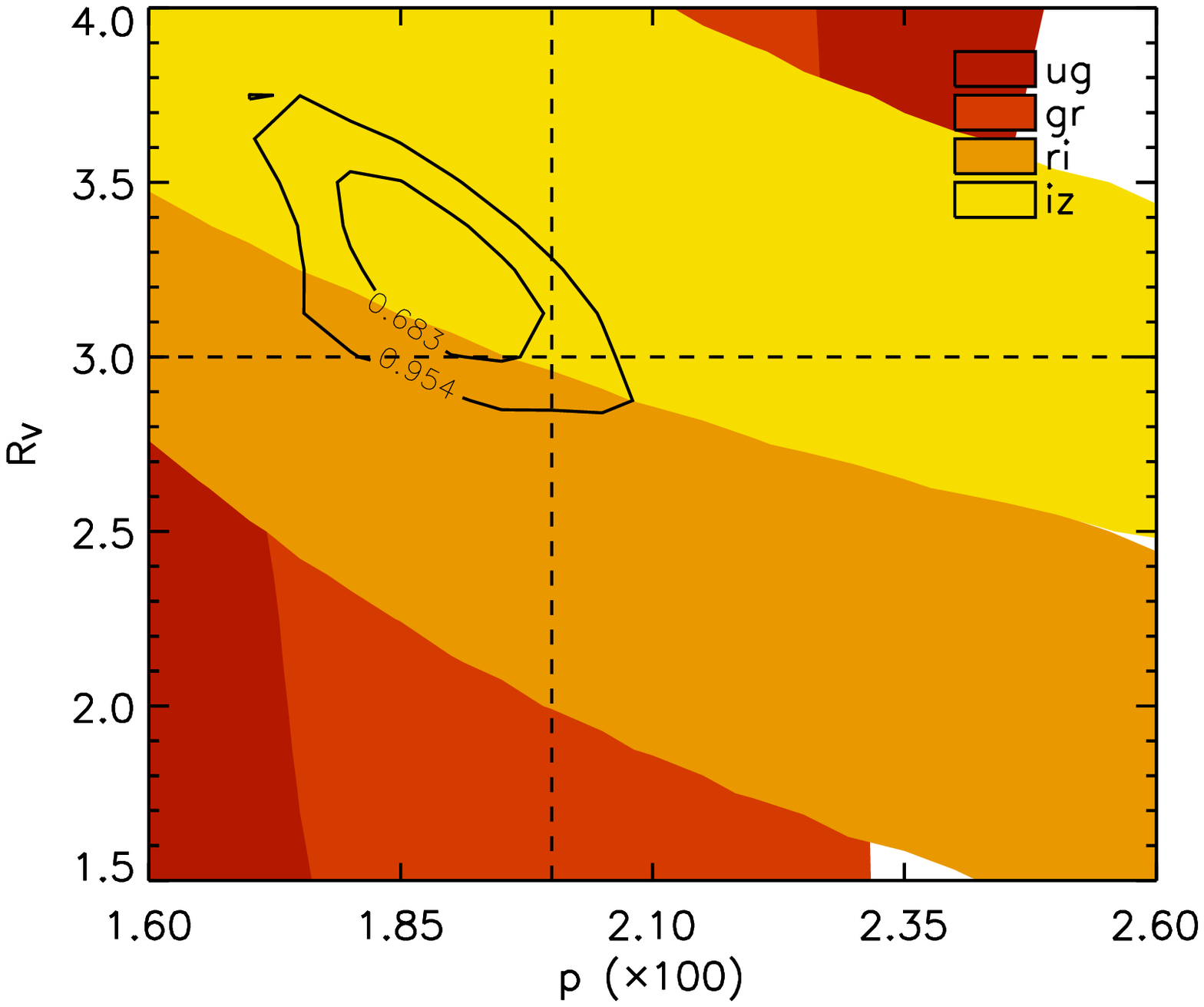}
\includegraphics[width=8cm]{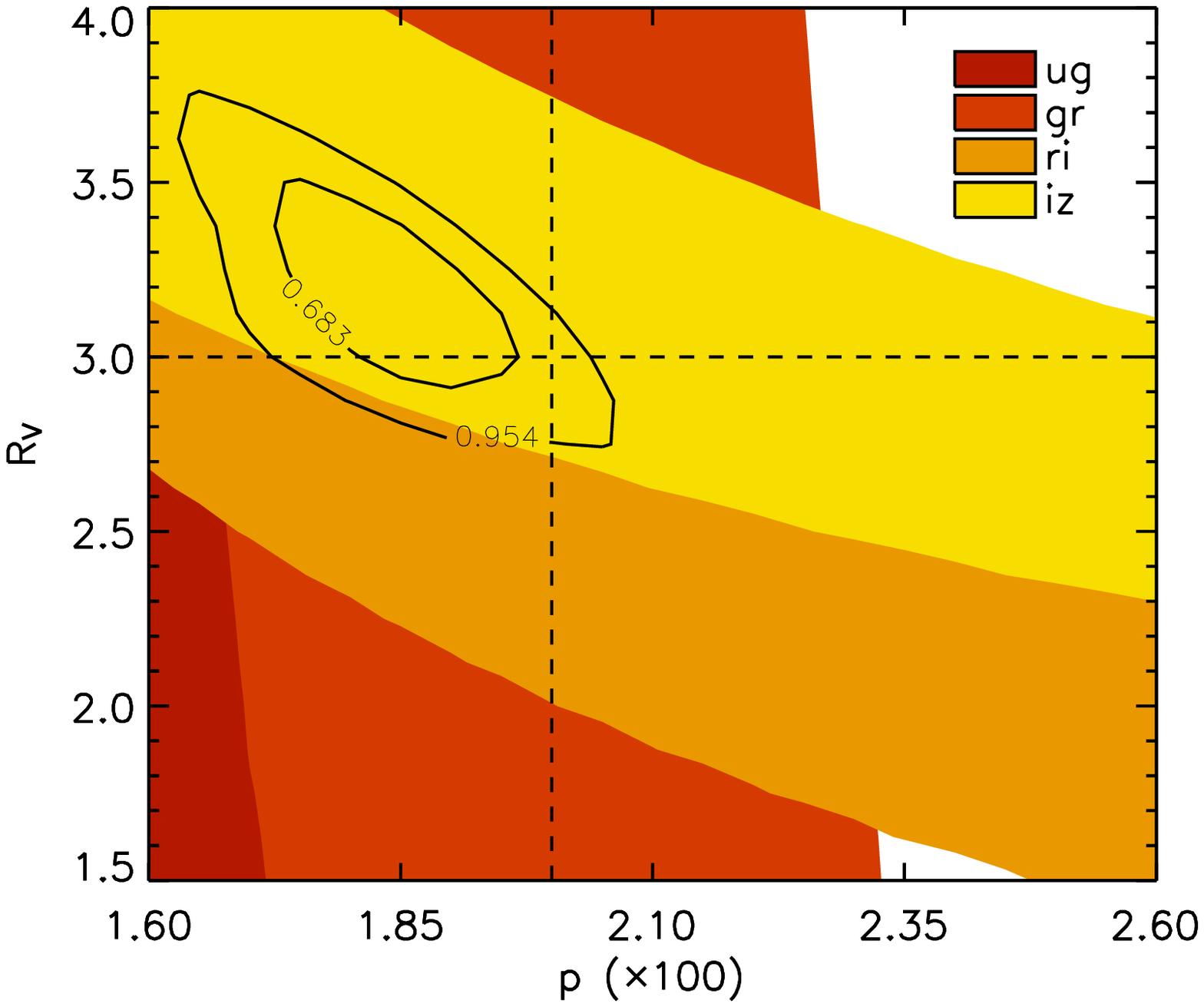}
\caption{\label{fig:combfitdc0-3} Constraints on $p$ and $R_V$ when
  dividing data into regions with low ($D^T<3$ MJy sr$^{-1}$) and high
  ($D^T>3$ MJy sr$^{-1}$) dust column density. There is no evidence
  for any correlation between the dust column density and dust
  properties.  {\em Upper left:} Results for CCM dust for all regions
  with $D^T<3$ MJy sr$^{-1}$ corresponding to 95.4\,\% CL for one
  parameter of $p=0.020 \pm 0.0016$ and $R_V=3.06 \pm 0.56$.  {\em
    Upper right:} Results for CCM dust at $D^T>3$ MJy sr$^{-1}$ giving
  $p=0.019 \pm 0.0019$ and $R_V=3.09 \pm 0.65$.  {\em Lower left:}
  Results for Fitzpatrick dust at $D^T<3$ MJy sr$^{-1}$ giving
  $p=0.019 \pm 0.0016$ and $R_V=3.28 \pm 0.37$.  {\em Lower right:}
  Results for Fitzpatrick dust at $D^T>3$ MJy sr$^{-1}$ giving
  $p=0.018 \pm 0.0018$ and $R_V=3.23 \pm 0.41$.}
\end{figure*}

It is not unreasonable to assume that the distributions of grain
sizes, and thus the resulting value of $R_V$, is different in high-
and low-density environments. Whether grains are generally larger or
smaller in high-density environments depends on whether grain-grain
collisions are more likely to cause coalescence or shattering.  We
therefore check for any evidence of correlations between $R_V$ and
the dust column density by dividing our sample into subsamples with
different column densities. Note, however, that a high column density
does not necessarily correspond to a high physical density.
In Fig.~\ref{fig:combfitdc0-3} we show constraints from dividing the QSO sample into
two regions with low ($D^T<3$ MJy sr$^{-1}$) and high ($D^T>3$ MJy sr$^{-1}$) dust 
column density.
Our constraints for CCM are $p=0.020 \pm 0.0016$ and $R_V=3.06 \pm 0.56$ ($D^T<3$ MJy sr$^{-1}$),
and $p=0.019 \pm 0.0019$ and $R_V=3.09 \pm 0.65$ ($D^T>3$ MJy sr$^{-1}$).
For Fitzpatrick dust, $p=0.019 \pm 0.0016$ and $R_V=3.28 \pm 0.37$ ($D^T<3$ MJy sr$^{-1}$),
and $p=0.018 \pm 0.0018$ and $R_V=3.23 \pm 0.41$ ($D^T>3$ MJy sr$^{-1}$)
(95.4\,\% CL for one parameter). This simple test does not provide evidence for 
any correlation between the dust column density and
dust properties.

\subsection{$\sigma(R_V)$ and $\sigma(A_V)$}\label{sec:sigrvandsigav}
\begin{figure*}[!htb]
\centering
\includegraphics[width=8cm]{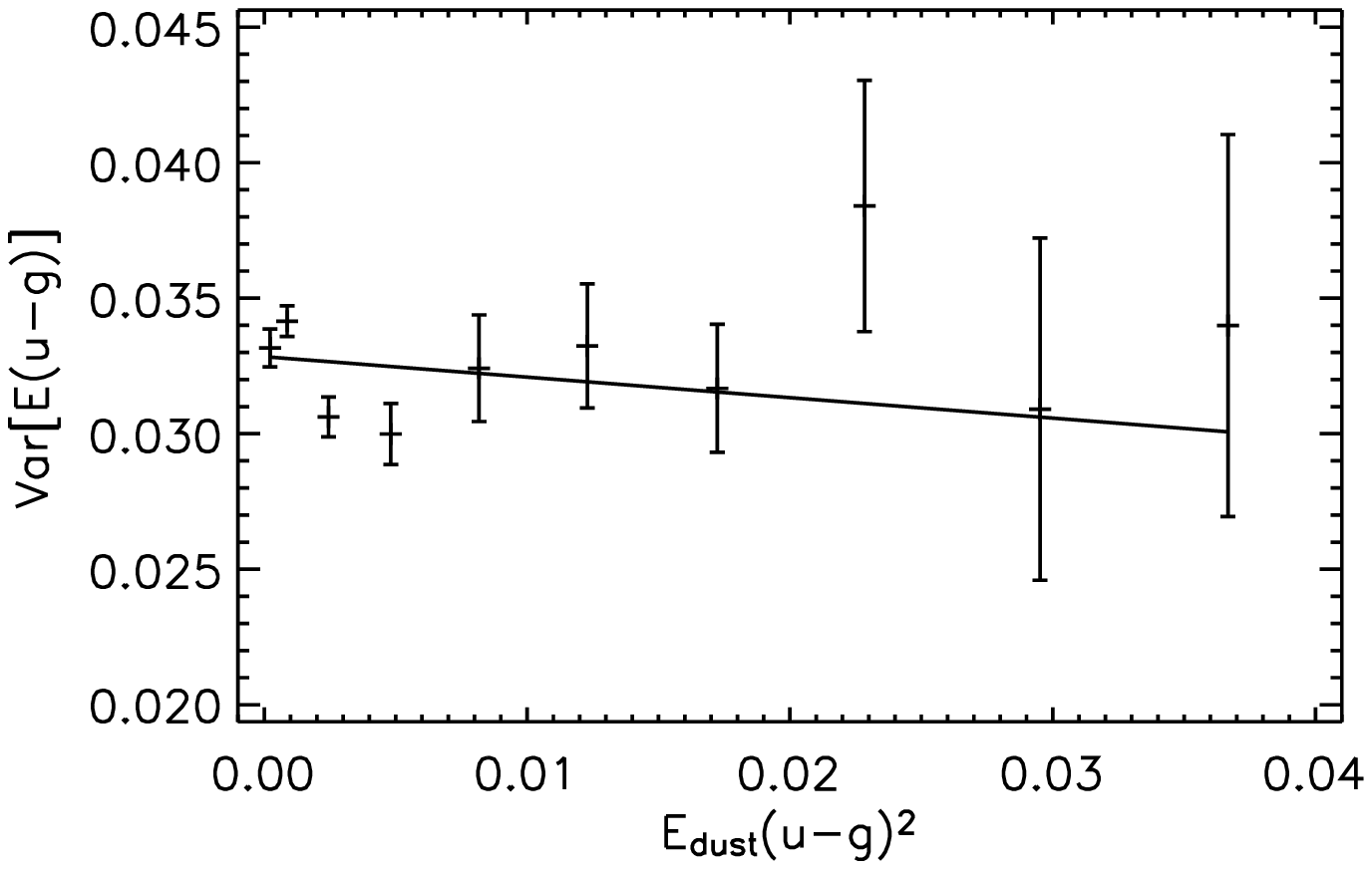}
\includegraphics[width=8cm]{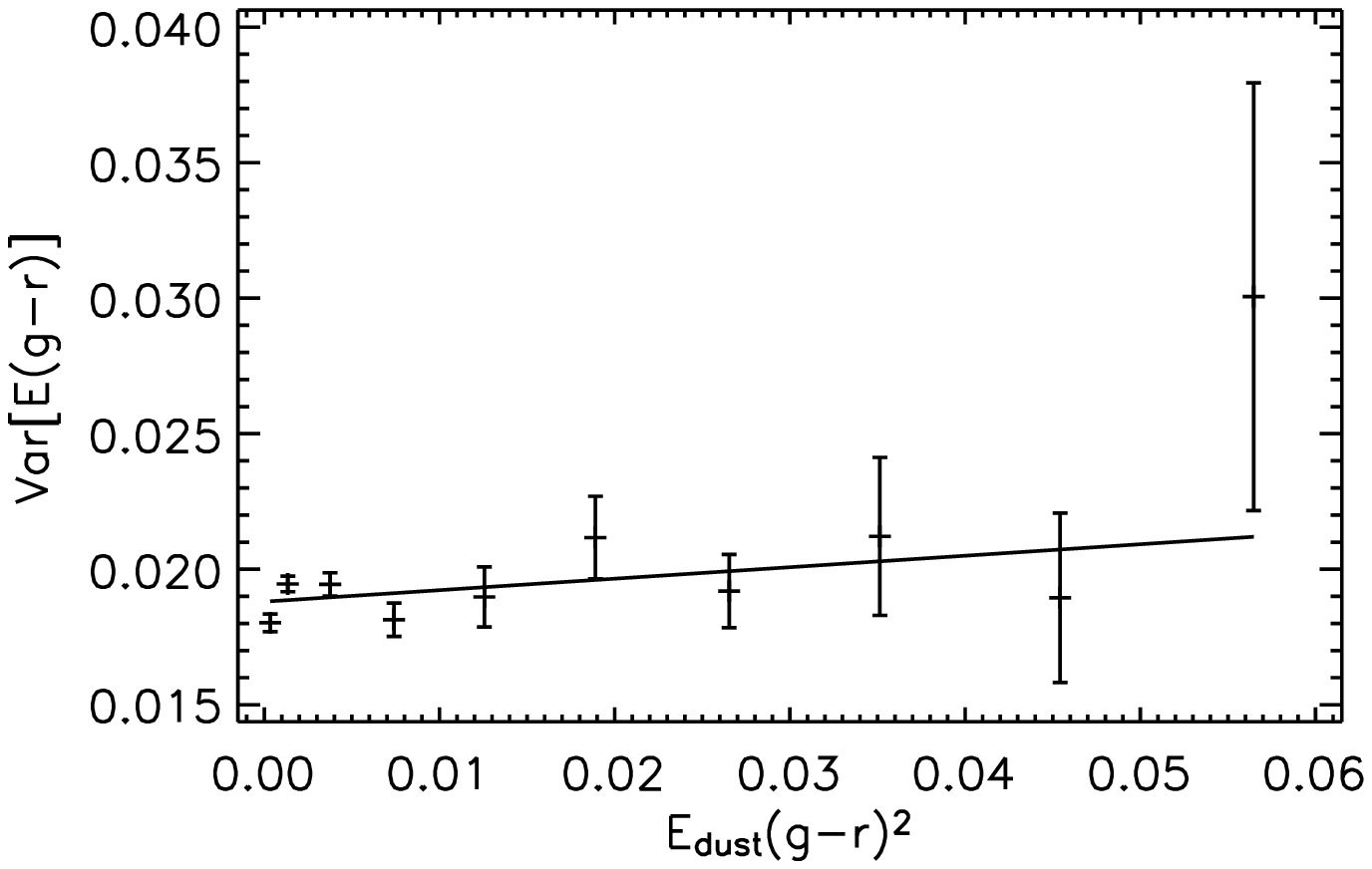}
\includegraphics[width=8cm]{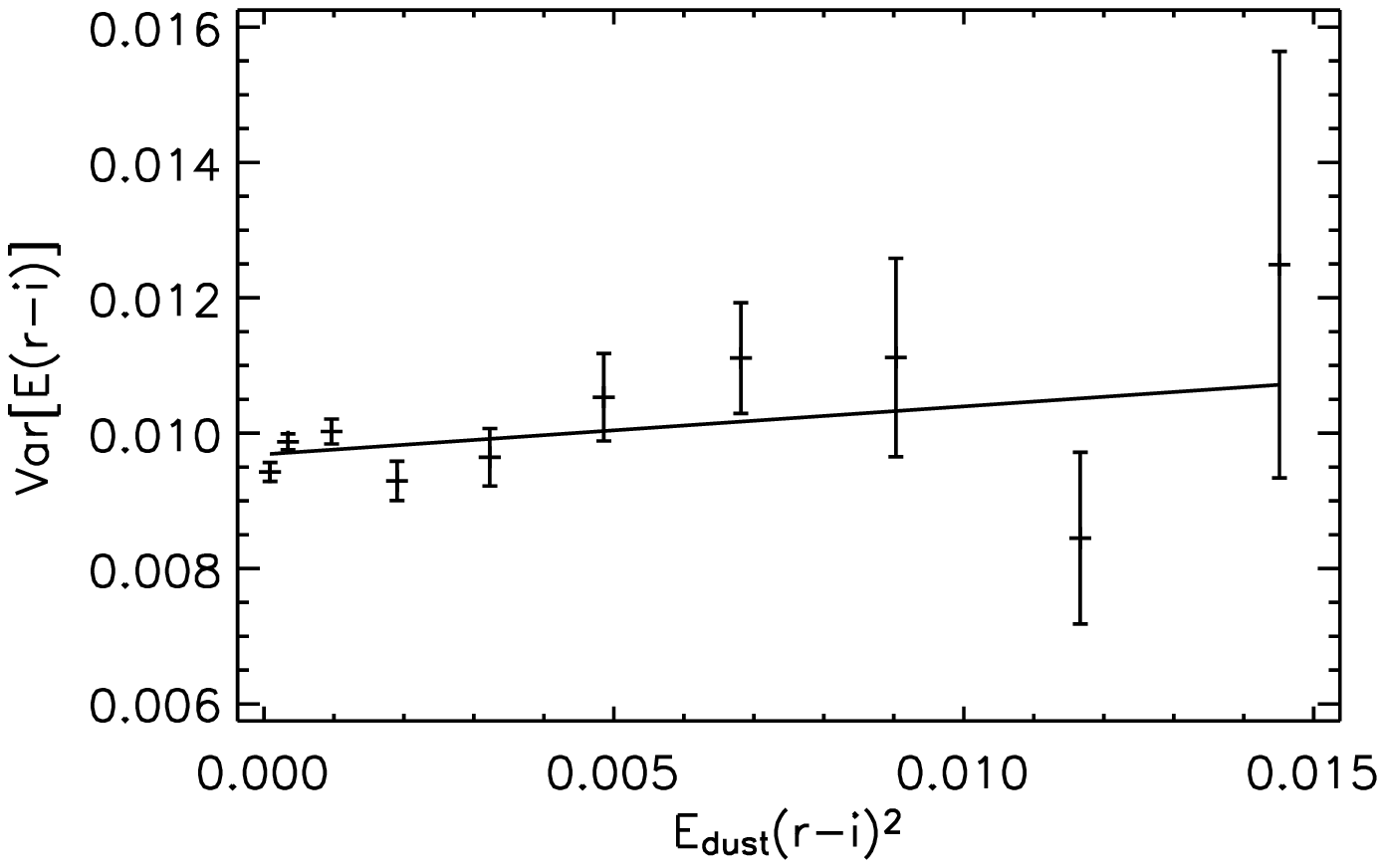}
\includegraphics[width=8cm]{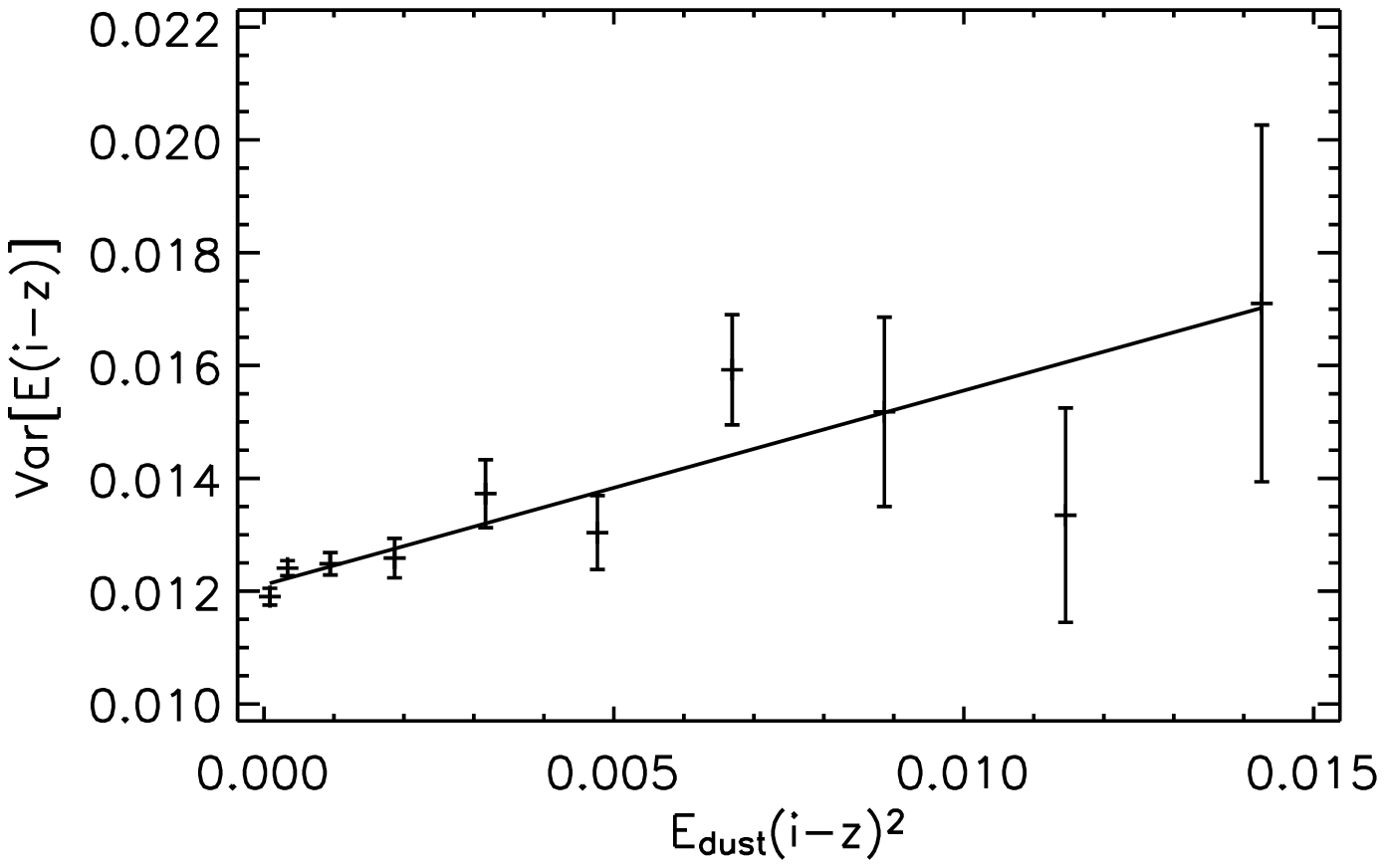}
\caption{\label{fig:colscat} The variance $\sigma[E(X-Y)]^2$ as a function of $E_d(X-Y)^2$ for
each standard colour of the QSO sample. The total variance at zero $E_d(X-Y)^2$ is the 
intrinsic colour variance $\sigma[E_i(X-Y)]^2$ and the slope is related to the scatter in $R_V$ and $A_V$.}
\end{figure*}

\begin{figure*}[!htb]
\centering
\includegraphics[width=8cm]{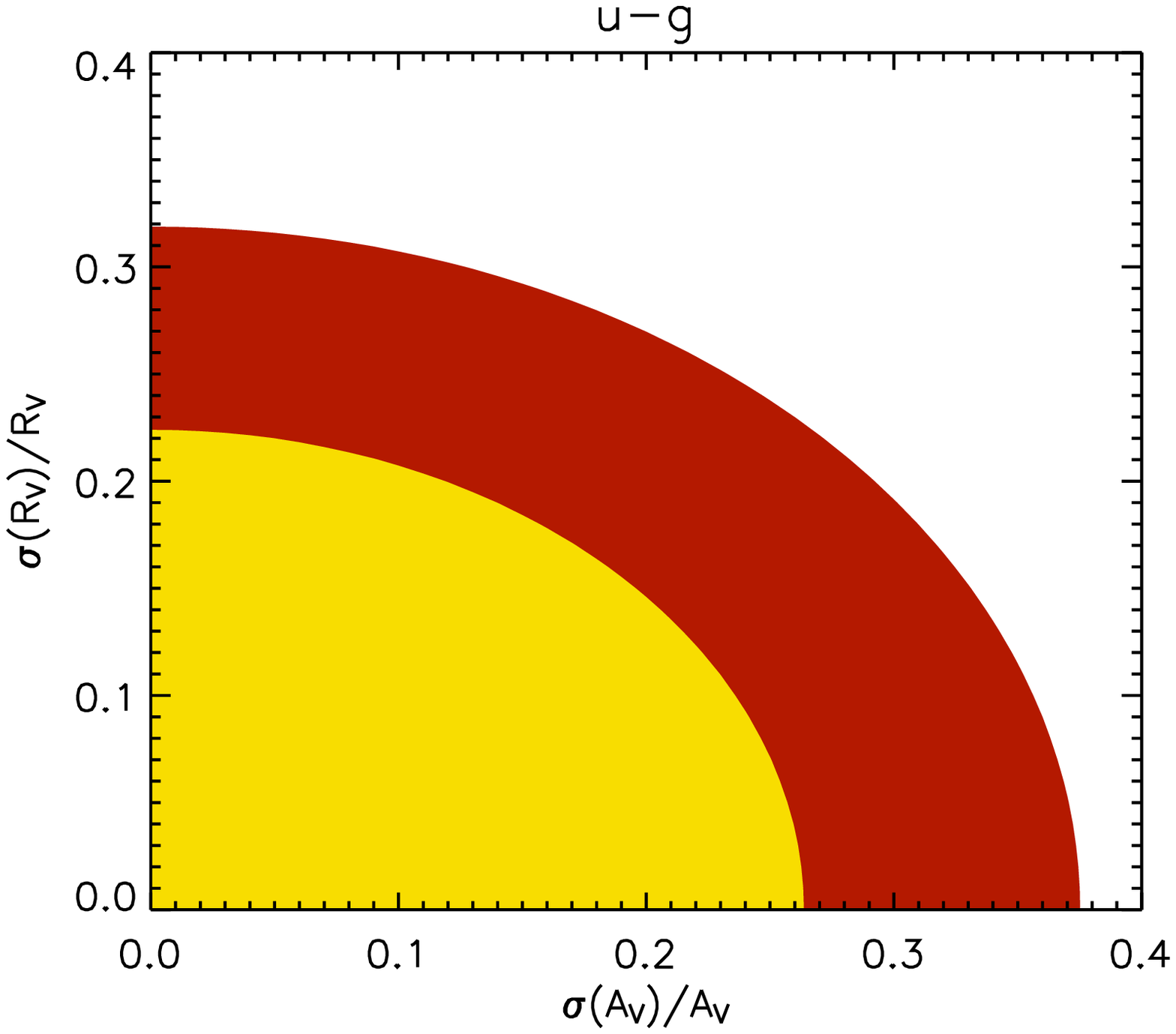}
\includegraphics[width=8cm]{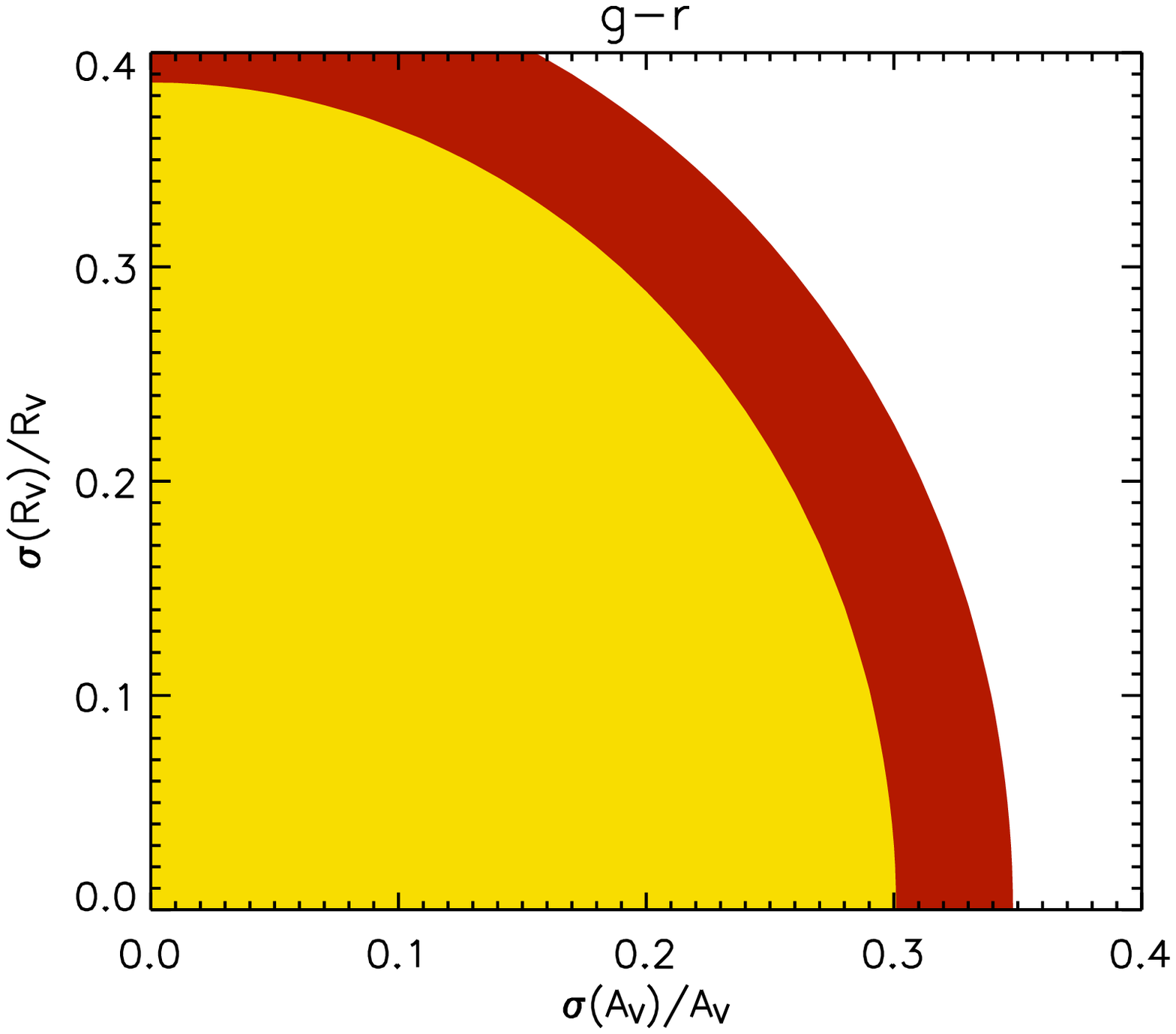}
\includegraphics[width=8cm]{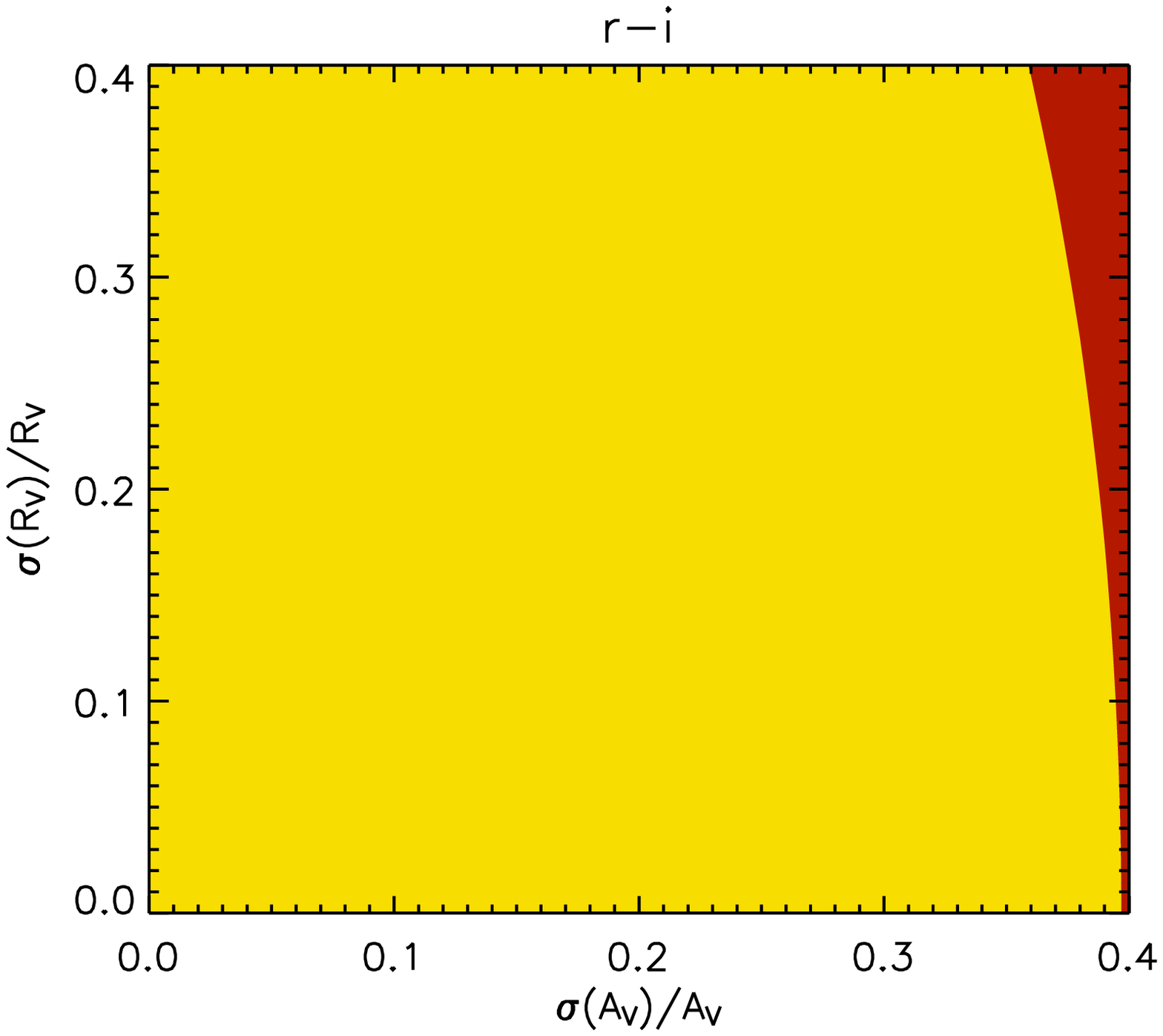}
\includegraphics[width=8cm]{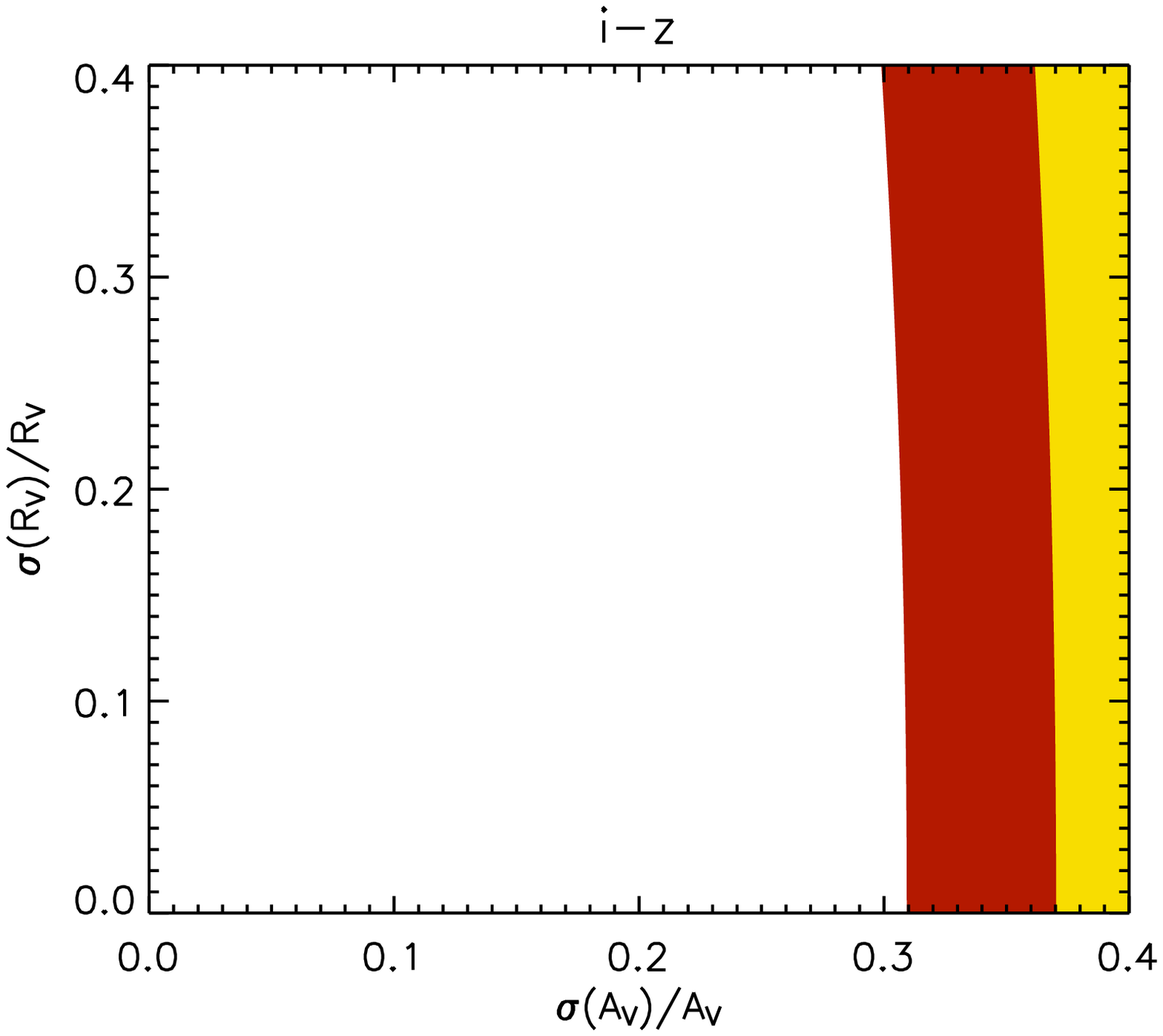}
\caption{\label{fig:colscatcon} Constraints on constant $\sigma_{R_V}/R_V$ and
$\sigma_{A_V}/A_V$ derived from the linear fits depicted in Fig.~\ref{fig:colscat}.
The coloured regions correspond to 95.4\,\% CL (red) and 68.3\,\% CL (yellow) for two parameters.}
\end{figure*}

If dust has spatially varying properties, depending on the size of the
regions we would expect the colour dispersion to be higher behind
regions with high column density $D^T$.
For filters $X$ and $Y$ we can write the colour $E(X-Y)$ as a
combination of the intrinsic colour $E_i(X-Y)$ and the dust induced
colour $E_d(X-Y)$
\begin{equation}
  E(X-Y)= E_i(X-Y)+ E_d(X-Y)\, ,
\end{equation}
and the corresponding colour dispersion
\begin{equation}
  \sigma[E(X-Y)]^2= \sigma[E_i(X-Y)]^2+ \sigma[E_d(X-Y)]^2\, ,
\end{equation}
where
\begin{equation}
  E_d(X-Y)= \left(\frac{A_X}{A_V}-\frac{A_Y}{A_V}\right)A_V\equiv f_{XY}(R_V)A_V\, , 	
\end{equation}
and
\begin{eqnarray}
  \sigma[E_d(X-Y)]^2
    &=& \left[\left(\frac{R_V}{f_{XY}}\frac{\partial f_{XY}}{\partial R_V}\right )^2\left(\frac{\sigma_{R_V}}{R_V}\right)^2+
      \left(\frac{\sigma_{A_V}}{A_V}\right)^2\right]E_d(X-Y)^2\\ \nonumber
      &\equiv& g[X,Y,R_V,\sigma(R_V),\sigma(A_V)])E_d(X-Y)^2\, .
\end{eqnarray}
In Fig.~\ref{fig:colscat}, we plot the variance $\sigma[E(X-Y)]^2$ as
a function of $E_d(X-Y)^2$ for each standard colour of the QSO sample,
expecting the variance at zero $E_d(X-Y)^2$ to give the intrinsic
colour variance $\sigma[E_i(X-Y)]^2$ and the slope to be related to
the scatter in $R_V$ and $A_V$ through the function
$g[X,Y,R_V,\sigma(R_V),\sigma(A_V)]$ (using the CCM dust extinction
parametrisation). As can be seen, there are no obvious correlations
between $\sigma[E(X-Y)]^2$ and $E_d(X-Y)^2$ (or $\sigma[E(X-Y)]^2$ and
the dust column density $D^T$). Assuming $\sigma_{R_V}/R_V$ and
$\sigma_{A_V}/A_V$ to be constant, although the results depicted in
Fig.~\ref{fig:colscat} indicate that this assumption may be too
simplistic, we can constrain their values from the fitted straight
line slopes. Results from this analysis are shown in
Fig.~\ref{fig:colscatcon}.  The bluer colours show no increase in the
colour variance with dust column density, whereas $i-z$ show a slight
increase in the variance indicating a possible dispersion of
$\sigma_{A_V}/A_V$ at the 30\,\% level. This said, there are a few
points that should be remembered: First, this is under the assumption
that the relation between the variance and the colour is indeed
linear, and that the data points are not correlated. Second, results
for $i-z$ are not consistent with the bluer colours that show no signs
of a scatter in the dust properties, although it is possible that the
scatter only manifests itself at longer wavelengths. Third, there is
a weak binning dependence when calculating the dispersion as a
function of $E_d(X-Y)^2$.  To summarize, there is no evidence in the
QSO data for any additional scatter in the observed colours in regions
with high column density, which indicates that dust properties are
relatively homogeneous averaged over the size of spatial regions
probed in this paper, except perhaps for a hint of slightly increased
scatter at the reddest optical wavelengths probed by the QSO data.

\section{Summary}\label{sec:summary}
By investigating the correlations between dust column density as
inferred from infrared data and the observed colours of celestial
objects at cosmological distances with low levels of colour dispersion, we
constrain the properties of Milky Way dust. We expect analyses
of the reddening of extragalactic sources to complement analyses of
Galactic sources very well since they will have different systematic
uncertainties, and in principle may also probe dust at different
distances. Results from QSOs, BCGs, and LRGs colours are broadly
consistent, indicating a proportionality constant between the
reddening $E(B-V)\equiv A_B-A_V$ and the dust column density $D^T$
(given in units of MJy/sr) of $p=E(B-V)/D^T\sim 0.02$ and a reddening
parameter $R_V\equiv A_V/E(B-V)\sim 3$ with relative uncertainties of
approximately $10\,\%$.  In spite of the increased statistics, the
fractional error of $p$ is similar to that of
\citet{1998ApJ...500..525S}, the reason being that we simultaneously
fit for $R_V$ and that the uncertainty is dominated by the possibility
of systematic shifts in the effective filter wavelengths.

We note that, in accordance with the analyses in
\citet{2010ApJ...725.1175S} and \citet{2011ApJ...737..103S}, the FTZ
extinction curve may be argued to be preferred before the CCM
parametrisation, since it is able to fit QSO, BCG, and LRG data with
similar normalisations and reddening parameters. However, as argued in
Sect.~\ref{sec:disc}, such differences may also be traced to
uncertainties in the spectral template functions.

The data does not provide evidence for any dependence of the dust
properties as a function of dust column density, nor for any
sizable scatter in the dust properties on the scales probed by the
data, except for a hint of scatter $\sigma_{A_V}/A_V\ga 0.3$ in the
extinction properties of dust at the longest wavelengths probed by the
data (corresponding to $i-z$).  We have also checked that within the
uncertainties there is no evidence of any systematic difference
between the dust extinction properties in the northern and southern
hemisphere.  Assuming the dust maps of \citet{1998ApJ...500..525S} to
be correct, it is fair to assume that we can correct for Milky
Way dust by applying a proportionality constant $p=E(B-V)/D^T\sim
0.02$ with reddening parameter $R_V\sim 3$ to the dust maps.

\bibliographystyle{aa}
\bibliography{paper}

\begin{acknowledgements}
EM acknowledges support for this study by the Swedish Research Council.
EM also acknowledges discussions with Brice Menard and Ariel Goobar.

Funding for the SDSS and SDSS-II has been provided by the Alfred
P. Sloan Foundation, the Participating Institutions, the National
Science Foundation, the US Department of Energy, the National
Aeronautics and Space Administration, the Japanese Monbukagakusho, the
Max Planck Society, and the Higher Education Funding Council for
England. The SDSS is managed by the Astrophysical Research Consortium
for the Participating Institutions. The Participating Institutions are
the American Museum of Natural History, Astrophysical Institute
Potsdam, University of Basel, Cambridge University, Case Western
Reserve University, University of Chicago, Drexel University,
Fermilab, the Institute for Advanced Study, the Japan Participation
Group, Johns Hopkins University, the Joint Institute for Nuclear
Astrophysics, the Kavli Institute for Particle Astrophysics and
Cosmology, the Korean Scientist Group, the Chinese Academy of Sciences
(LAMOST), Los Alamos National Laboratory, the Max Planck Institute for
Astronomy (MPIA), the Max Planck Institute for Astrophysics (MPA), New
Mexico State University, The Ohio State University, University of
Pittsburgh, University of Portsmouth, Princeton University, the United
States Naval Observatory, and the University of Washington.
\end{acknowledgements}


\end{document}